\documentclass[10pt]{IEEEtran}

\makeatletter
\def\ps@headings{%
\def\@oddhead{\mbox{}\scriptsize\rightmark \hfil \thepage}%
\def\@evenhead{\scriptsize\thepage \hfil \leftmark\mbox{}}%
\def\@oddfoot{}%
\def\@evenfoot{}}
\makeatother 

\usepackage{epstopdf}
\usepackage{graphicx}
\usepackage{stmaryrd}
\usepackage{amssymb,amsmath}
\usepackage{multirow,url}
\usepackage{color,cite}

\newtheorem{proposition}{Proposition}
\newtheorem{definition}{Definition}
\newtheorem{theorem}{Theorem}
\newtheorem{observation}{Observation}
\newtheorem{game}{Game}

\ifodd 0
\newcommand{\rev}[1]{{\color{blue}#1}} 
\newcommand{\com}[1]{\textbf{\color{red} (COMMENT: #1)}} 
\newcommand{\comg}[1]{\textbf{\color{green} (COMMENT: #1)}}
\newcommand{\response}[1]{\textbf{\color{magenta} (RESPONSE: #1)}} 
\else
\newcommand{\rev}[1]{#1}
\newcommand{\com}[1]{}
\newcommand{\comg}[1]{}
\newcommand{\response}[1]{}
\fi

\begin{document}

\title{Competition with Dynamic Spectrum Leasing}

\author{Lingjie Duan$^{1}$, Jianwei Huang$^{1}$, and Biying Shou$^{2}$\\
$^{1}$Department of Information
Engineering, the Chinese University of Hong Kong, Hong Kong \\
$^{2}$Department of Management Sciences, City University of Hong
Kong, Hong Kong
\thanks{Part of the results will appear in the \emph{IEEE Symposium on International Dynamic Spectrum Access Networks (DySPAN)}, Singapore, April 2010 \cite{DuanHuangShou09b}.}}

\maketitle

\begin{abstract}
This paper presents a comprehensive analytical study of two
competitive cognitive operators' spectrum leasing and pricing
strategies, taking into account operators' heterogeneity in leasing
costs and users' heterogeneity in transmission power and channel
conditions.
%
We model the interactions between operators and users as a
three-stage dynamic game, where operators make simultaneous spectrum
leasing and pricing decisions in Stages I and II, and users make
purchase decisions in Stage III.
Using backward induction, we are able to completely characterize the
game's equilibria. We show that both operators make the equilibrium
leasing and pricing decisions based on simple threshold policies.
Moreover, two operators always choose the same equilibrium price
despite their difference in leasing costs. Each user receives the
same signal-to-noise-ratio (SNR) at the equilibrium, and the
obtained payoff is linear in its transmission power and channel
gain.
We also compare the duopoly equilibrium with the coordinated case
where two operators cooperate to maximize their total profit. We
show that the maximum loss of total profit due to operators'
competition is no larger than $25\%$. The users, however, always
benefit from operators' competition in terms of their payoffs.
\rev{We show that most of these insights are robust in the general SNR
regime.} \com{Though we have made the high SNR assumption in order
to obtain closed-form equilibrium results, most of the engineering
insights remain valid in the general SNR regime.}
\end{abstract}

\section{Introduction}


Wireless spectrum is often considered as a scarce resource, and thus
has been tightly controlled by the governments through static
license-based allocations. However, several recent field
measurements show that many spectrum bands are often under-utilized
even in densely populated urban areas \cite{sharedspectrum2005}. To
achieve more efficient spectrum utilization, various dynamic
spectrum access mechanisms have been proposed so that
unlicensed secondary users can share the spectrum with the
licensed primary users.
One of the proposed mechanisms is dynamic spectrum leasing, where a
spectrum owner dynamically transfers and trades the usage right of
temporarily unused part of its licensed spectrum to secondary
network operators or users in exchange of monetary compensation
(e.g., \cite{jayaweera2009dynamic,IEEEhowto:Zhao,
Simeone,IEEEhowto:Jia,Chapin}). In this paper, we study the competition of two secondary operators under the dynamic spectrum leasing mechanism.



Our study is motivated by the successful operations of mobile
virtual network operators (MVNOs) in many countries
today\footnote{There are over 400 mobile virtual network operators
owned by over 360 companies worldwide as of February 2009
\cite{wiki:MVNO}.}. An MVNO does not own wireless spectrum
or even the physical infrastructure. It provides services to
end-users by \emph{long-term} spectrum leasing agreements with a
spectrum owner. MVNOs are similar to the ``switchless resellers'' of
the traditional landline telephone market. Switchless resellers buy
minutes wholesale from the large long distance companies and resell
them to their customers. As intermediaries between spectrum owners
and users, MVNOs can raise the competition level of the wireless
markets by providing competitive pricing plans as well as more
flexible and innovative value-added services. However, an MVNO is
often stuck in a long-term contract with a spectrum owner and can
not make flexible spectrum leasing and pricing decisions to match
the dynamic demands of the users.
The secondary operators considered in this paper do not own wireless spectrum either.
Compared with a traditional MVNO, the secondary operators can dynamically adjust their spectrum leasing and
pricing decisions to match the users' demands that change with
users' channel conditions.


This paper studies the competition of two secondary operators (also
called \emph{duopoly}) who compete  to serve a
common pool of secondary users. The secondary operators will
dynamically lease spectrum from spectrum owners, and then compete to
sell the resource to the secondary users to maximize their
individual profits. We would like to understand \emph{how the
operators make the equilibrium investment (leasing) and pricing
(selling) decisions, \rev{considering operators' heterogeneity in
leasing costs and wireless users' heterogeneity in transmission
power and channel conditions}\com{when taking the details of the
wireless technology into consideration}.}

We adopt a three-stage dynamic game model to study the (secondary)
operators' investment and pricing decisions as well as the
interactions between the operators and the (secondary) users. From
here on, we will simply use ``operator'' to denote ``secondary
operator'' and ``users'' to denote ``secondary users''. In Stage I,
the two operators simultaneously lease spectrum (bandwidth) from the
spectrum owners with different leasing costs. In Stage II, the two
operators simultaneously announce their spectrum retail prices to
the users. In Stage III, each user determines how much resource to
purchase from which operator. Each operator wants to maximizes its
profit, which is the difference between the revenue collected from
the users and the cost paid to the spectrum owner.





Key results and contributions of this paper include:

\begin{itemize}
\item{\emph{A concrete wireless spectrum sharing model}}:  We assume that   users share the spectrum using orthogonal frequency
division multiplexing (OFDM) technology. A user's achievable rate depends on its allocated bandwidth, maximum transmission power, and channel condition. This model is more concrete than several generic
economic models used in related literature (e.g.,
\cite{IEEEhowto:Jia,IEEEhowto:Niyato,IEEEhowto:Xing,IEEEhowto:Ileri}),
and can provide more insights on how the wireless technology impact the operators' equilibrium economic decisions.

\item \emph{Symmetric pricing equilibrium}: We show the two operators' always choose the same equilibrium price, even when they have different leasing costs and make different equilibrium investment decisions. Moreover, this price is independent of users' transmission power and channel conditions.

\item \emph{Threshold structures of investment and pricing equilibrium}:
We show that the operators' equilibrium investment and pricing
decisions follow simple threshold structures, which are easy to
implement in practice.

\item \emph{Fair Quality of Service (QoS) of users}:
We show that each user achieves the same signal-to-noise (SNR) that is independent of the users' population and wireless characteristics.


\item \emph{Impact of competition}:
We show that the operators' competition  leads to a
maximum loss of 25\%  in terms of the two operators' total profit compared with a coordinated case. The users, however, always benefit from the operators' competition by achieving better payoffs.

%
\end{itemize}

Next we briefly discuss the related literature. In Section
\ref{sec:NetworkModel}, we describe the network model and game
formulation. In Section \ref{sec:BackwardInduction}, we analyze the
dynamic game through backward induction and calculate the duopoly
leasing/pricing equilibrium. We discuss various insights obtained
from the equilibrium analysis in Section \ref{sec:Equilibrium}. In
Section \ref{sec:Efficiency}, we show the impact of duopoly
competition on the total operators' profit and the users' payoffs.
We conclude in Section \ref{sec:conclusion} together with some
future research directions.

\subsection{Related Work}
The existing results on dynamic spectrum access mainly focused on
the technical aspects of primary users' spectrum sharing with
secondary users. Two approaches are extensively studied: (1)
spectrum underlay, which allows secondary users to coexist with
primary users by imposing constraints on the transmission powers of
secondary users (e.g.,
\cite{IEEEhowto:Etkin,RMenon,SHuang,Scutari2009}); (2) spectrum
overlay, which allows secondary users to identify and exploit
spatial and temporal spectrum availability in a nonintrusive manner
(e.g.,
\cite{Papadimitratos,QingZhao,ZhaoTong,huang2009optimal,HuangLiuDing2009}).
These results did not consider the spectrum owners' economic
incentive in sharing spectrum with secondary users.

Recently researchers started to study the economic aspect of dynamic spectrum access, such as the cognitive secondary operators' strategies of spectrum acquisition from spectrum owners and service provision to the users. For example, several auction mechanisms have been proposed for the spectrum owner to allocate spectrum (e.g., \cite{IEEEhowto:Huang,IEEEhowto:Niyato,jia2009revenue,Gandhi,Rodriguez,WangXu2010,WangLiXu2010}). Cognitive radio operators can also
obtain spectrum by dynamically leasing from the spectrum
owner (e.g., \cite{jayaweera2009dynamic,IEEEhowto:Jia,Chapin,Simeone,report_Infocom}).

For operators' service provision, most related results looked at the
pricing interactions between cognitive network operators and the
secondary users (e.g.,
\cite{sengupta2009economic,IEEEhowto:Jia,IEEEhowto:Niyato,IEEEhowto:Ileri,IEEEhowto:Xing,Niyato2}).
Ref.~\cite{IEEEhowto:Jia} and \cite{IEEEhowto:Niyato} studied the
pricing competition among two or more operators.
Ref.~\cite{IEEEhowto:Xing} explored users' demand functions in both
quality-sensitive and price-sensitive buyer population models.
Ref.~\cite{IEEEhowto:Ileri} derived users' demand functions based on
the acceptance probability model. Ref.~\cite{sengupta2009economic}
considered users' queuing delay due to congestion in spectrum
sharing. Ref.~\cite{Niyato2} modeled the dynamic behavior of
secondary users as an evolutionary game, and proposed an iterative
algorithm for operators' strategy adaption. Among these works, only
\cite{sengupta2009economic, Niyato2} studied practical wireless
spectrum sharing models by modeling users' wireless details. Many
results have been obtained mainly through extensive simulations
(e.g.,
\cite{sengupta2009economic,Niyato2,IEEEhowto:Niyato,IEEEhowto:Ileri,IEEEhowto:Xing}).

This work are related to our previous study\cite{report_Infocom},
where we considered the optimal sensing and leasing decisions of a
\emph{single} secondary operator facing supply uncertainty. The
focus of this paper is to study the competition between two
operators.


Another closely related paper is \cite{IEEEhowto:Jia}, which also jointly considered the the spectrum acquisition and service provision for cognitive operators. The key difference here is that we
present a comprehensive  analytical study that
characterizes the duopoly equilibrium investment and pricing
decisions, with \emph{heterogeneous} leasing costs for the operators and a
\emph{concrete} wireless spectrum sharing model for the users.

\section{Network Model}\label{sec:NetworkModel}

\begin{figure}[tt]
\centering
\includegraphics[width=0.4\textwidth]{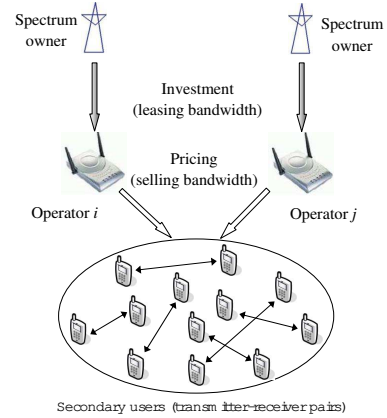}
\caption{Network model for the cognitive network operators.}
\label{fig:network}
\end{figure}

We consider two operators ($i,j\in\{1,2\}$ and {$i\neq j$}) and a
set $\mathcal{K}=\{1,\ldots,K\}$ of users as shown in Fig.
\ref{fig:network}. The operators obtain wireless spectrum from
different spectrum owners with different leasing costs, and compete
to serve the same set $\mathcal{K}$ of users. Each user has a
transmitter-receiver pair.
We assume that users are equipped with software defined radios and
can transmit in a wide range of frequencies as instructed by the
operators, but do not have the cognitive learning capacity. Such a
network structure puts most of the implementation complexity for
dynamic spectrum leasing and allocation on the operators, and thus
is easier to implement than a ``full'' cognitive radio network
especially for a large number of users. A user may switch among
different operators' services (e.g. WiMAX, 3G) depending on
operators' prices. It is important to study the competition among
multiple operators as operators are normally not cooperative.

\begin{figure}[tt]
\centering
\includegraphics[width=0.45\textwidth]{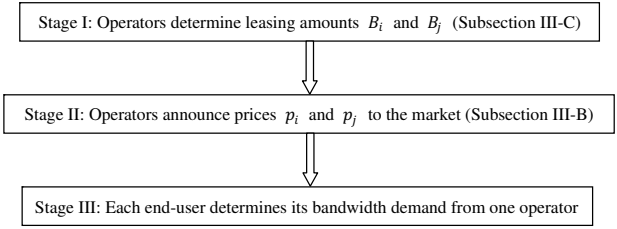}
\caption{Three-stage dynamic game: the duopoly's leasing and
pricing, and the users' resource allocation} \label{fig:threestage}
\end{figure}

The interactions between the two operators and the
users can be modeled as a \emph{three-stage dynamic game}, as shown
in Fig.~\ref{fig:threestage}. Operators $i$ and $j$ first
simultaneously determine their leasing bandwidths in Stage I, and then
simultaneously announce the prices to the users in Stage II.
Finally, each user chooses to purchase bandwidth from \emph{only one
operator} to maximize its payoff in Stage III.



%

The key notations of the paper are listed in Table \ref{tab:notation}. Some are explained as follows.
\begin{itemize}
\item \emph{Leasing decisions $B_{i}$ and $B_{j}$}: leasing bandwidths of operators $i$ and $j$ in Stage I, respectively.

\item \emph{Costs $C_{i}$ and $C_{j}$}: the fixed positive leasing costs per
unit bandwidth for operators $i$ and $j$, respectively. These costs
are determined by the negotiation between the operators and their
own spectrum suppliers.

\item \emph{Pricing decisions $p_i$ and $p_j$}: prices per unit bandwidth charged by operators $i$ and $j$ to the users in Stage II, respectively.

\item \emph{The User $k$'s demand $w_{ki}$ or $w_{kj}$}: the bandwidth demand of a user
$k\in\mathcal{K}$ from operator $i$ or $j$. A user can only purchase
bandwidth from one operator.

\end{itemize}

\begin{table}[tt]
\centering \caption{Key Notations}
\begin{tabular}{|p{0.9in}|p{2.3in}|}
\hline Notations & Physical Meaning \\\hline \hline $B_{i},B_{j}$ &
Leasing bandwidths of operators $i$ and $j$ \\\hline $C_{i},C_{j}$ &
Costs per unit bandwidth paid by operators $i$ and $j$\\\hline
$p_{i},p_{j}$ & Prices per unit bandwidth announced by operators $i$
and $j$\\\hline $\mathcal{K}=\{1,\ldots,K\}$ & Set of the users in
the cognitive network
\\\hline $P_k^{\max}$ & User $k$'s maximum
transmission power \\\hline $h_{k}$ & User $k$'s channel gain
between its transceiver
\\\hline $n_0$ & Noise power per unit bandwidth \\\hline $g_{k}=P_{k}^{\max}h_{k}/n_{0}$ &
User $k$'s wireless characteristic\\\hline
$G=\sum_{k\in\mathcal{K}}g_k$ & The users' aggregate wireless
characteristics\\\hline $w_{ki}, w_{kj}$ & User $k$'s bandwidth
allocation from operator $i$ or $j$\\\hline $r_k$ & User $k$'s data
rate
\\\hline $\mathcal{K}_i^P,\mathcal{K}_j^P$ & Preferred user sets of
operators $i$ and $j$\\\hline $D_i,D_j$ & Preferred demands of
operators $i$ and $j$\\\hline $\mathcal{K}_i^R,\mathcal{K}_j^R$ &
Realized user sets of operators $i$ and $j$\\\hline $Q_i,Q_j$ &
Realized demands of operators $i$ and $j$\\\hline $R_i,R_j$ &
Revenues of operators $i$ and $j$\\\hline$\pi_i,\pi_j$ & Profits of
operators $i$ and $j$\\\hline $T_{\pi}$ & Total profit of both
operators\\\hline
\end{tabular}
\tabcolsep 5mm \label{tab:notation}
\end{table}

\section{Backward Induction of the Three-Stage Game}
\label{sec:BackwardInduction}

A common approach of analyzing dynamic game is backward induction
\cite{Myerson}. We start with Stage III and analyze the users' behaviors
given the operators' investment and pricing decisions. Then we look
at Stage II and analyze how operators make the pricing decisions
taking the users' demands in Stage III into consideration. Finally, we look at the
operators' leasing decisions in Stage I knowing the results in
Stages II and III. Throughout the paper, we will use ``bandwidth'', ``spectrum'', and ``resource'' interchangeably.

\subsection{Spectrum Allocation in Stage III}\label{subsec:stageIII}
In Stage III, each user needs to make the following two decisions
based on the prices $p_i$ and $p_j$ announced by the operators in
Stage II:
\begin{enumerate}
\item Which operator to choose?
\item How much to purchase?
\end{enumerate}


\rev{OFDM has been deemed appropriate for dynamic spectrum sharing
(e.g., \cite{TAWeiss,Mahmoud})}. We assume that the users share the
spectrum using OFDM to avoid mutual interferences.\com{Lingjie: can
you find several recent good references that use OFDM for spectrum
sharing?} If a user $k\in\mathcal{K}$ obtains bandwidth $w_{ki}$
from operator $i$, then it achieves a data rate (in nats) of

%
%
\begin{equation}\label{eq:rate}
 r_k(w_{ki})= w_{ki}\ln\left(1+\frac{P_k^{\max}h_k}{n_0w_{ki}}\right),
\end{equation}
where $P_k^{\max}$ is user $k$'s maximum transmission power, $n_0$
is the noise power density, $h_k$ is the channel gain between user
$k$'s transmitter and receiver\cite{IEEEhowto:Junjik}. The channel
gain $h_{k}$ is independent of the operator, as the operator only
sells bandwidth and does not provide a physical
infrastructure.\footnote{We also assume that the channel condition
is independent of transmission frequencies, such as in the current
802.11d/e standard\cite{802} where the channels are formed by
interleaving over the tones. As a result, each user experiences a
flat fading over the entire spectrum.} Here we assume that user $k$
spreads its power $P^{\max}_{k}$ across the entire allocated
bandwidth $w_{ki}$.
To simplify later
discussions, we let
\begin{equation*}
g_k = P_k^{\max}h_k/n_0,
\end{equation*}
thus $g_k/w_{ki}$ is the user $k$'s SNR. The rate in (\ref{eq:rate})
is calculated based on the Shannon capacity.

To better obtain insights through closed-form solutions, we first
focus on the high SNR regime where $\mathtt{SNR}\gg1$. This will be
the case where a user  has limited choices of modulation and coding
schemes, and thus can not decode a transmission if the SNR is below
some threshold. In the high SNR regime, the rate in (\ref{eq:rate})
can be approximated as
\begin{equation}\label{eq:ratehighSNR}r_k(w_{ki})= w_{ki}
\ln\left(\frac{g_k}{w_{ki}}\right).\end{equation}
Although the analytical solutions in Section
\ref{sec:BackwardInduction} are derived based on
(\ref{eq:ratehighSNR}), we will show later in Section
\ref{sec:robust} that \emph{all major engineering insights remain
true in the general SNR regime.}

If a user $k$ purchases bandwidth $w_{ki}$ from operator $i$, it
receives a \emph{payoff} of
\begin{equation}\label{eq:utility}
u_k(p_i,w_{ki}) = w_{ki} \ln\left(\frac{g_k}{w_{ki}}\right)-p_i
w_{ki},
\end{equation}
which is the difference between the data rate and the payment. The
payment is proportional to price $p_{i}$ announced by operator $i$.
Payoff $u_k(p_i,w_{ki})$ is concave in $w_{ki}$, and the unique
\emph{demand} that maximizes the payoff is
\begin{equation}\label{eq:optbandwidth}
w_{ki}^*(p_i)=\arg\max_{w_{ki}\geq
0}u_{k}(p_i,w_{ki})=g_{k}e^{-(1+p_i)}.
\end{equation}
Demand $w_{ki}^*(p_i)$ is always positive, linear in $g_k$, and
decreasing in price $p_i$. Since $g_k$
is linear in channel gain $h_k$ and transmission power $P_k^{\max}$,
then a user with a better channel condition or a larger transmission
power has a larger demand. It is clear that $w_{ki}^*(p_i)$ is
upper-bounded by $g_k e^{-1}$ for any choice of price $p_i\geq 0$.
In other words, even if operator $i$ announces a zero price, user
$k$ will not purchase infinite amount of resource since it can not
decode the transmission if $\mathtt{SNR}_k=g_k/w_{ki}$ is low.

Eqn (\ref{eq:optbandwidth}) shows that every user purchasing
bandwidth from operator $i$ obtains the same SNR
$$\mathtt{SNR}_{k}=\frac{g_k}{w_{ki}^*(p_i)}=e^{(1+p_i)},$$
and obtains a payoff linear in $g_k$
$$u_{k}(p_i,w_{ki}^\ast(p_i))=g_k e^{-(1+p_i)}.$$

\subsubsection{Which Operator to Choose?} Next we explain how
each user decides which operator to purchase from.
The following definitions help the discussions.
\begin{definition}[Preferred User Set]
The Preferred User Set $\mathcal{K}_i^P$ includes the users who
prefer to purchase from operator $i$.
\end{definition}
\begin{definition}[Preferred Demand]
The Preferred Demand $D_i$ is the total demand from users in the
preferred user set $\mathcal{K}_i^P$, i.e.,
\begin{equation}\label{eq:demand1}
D_{i}(p_{i},p_{j})=\sum_{k\in\mathcal{K}_i^P(p_{i},p_{j})}g_{k}e^{-(1+p_i)}.
\end{equation}
\end{definition}

The notations in (\ref{eq:demand1}) imply that both set
$\mathcal{K}_i^P$ and demand $D_{i}$ only depend on prices
$(p_{i},p_{j})$ in Stage II and are independent of operators' leasing decisions $(B_{i},B_{j})$ in Stage I. Such dependance can be discussed in two cases:
%
%
\begin{enumerate}
\item \emph{Different Prices} ($p_i<p_j$): every user $k\in\mathcal{K}$ \emph{prefers} to purchase from
operator $i$ since
$$u_k(p_i,w_{ki}^{*}(p_i))>u_k(p_j,w_{kj}^{*}(p_j)).$$
We have $\mathcal{K}_i^P=\mathcal{K}$ and  $\mathcal{K}_j^P=\emptyset$, and
$$D_{i}(p_{i},p_{j})=Ge^{-(1+p_i)} \textrm{ and } D_{j}(p_{i},p_{j})=0,$$
where $G=\sum_{k\in\mathcal{K}}g_{k}$ represents the aggregate wireless characteristics of the users. This notation will be used heavily later in the paper.

\item \emph{Same Prices} ($p_i=p_j=p$): every user $k\in\mathcal{K}$ is indifferent between the operators and randomly chooses one with equal probability. In this case,
$$D_{i}(p,p)=D_{j}(p,p)=Ge^{-(1+p)}/2.$$
\end{enumerate}

Now let us discuss how much demand an operator can actually satisfy,
which depends on the bandwidth investment decisions $(B_{i},B_{j})$ in Stage I.
It is useful to define the following
terms.
\begin{definition}[Realized User Set]
The Realized User Set $\mathcal{K}_i^R$ includes the users whose
demands are satisfied by operator $i$.
\end{definition}
\begin{definition}[Realized Demand]
The Realized Demand $Q_i$ is the total demand of users in the
Realized User Set $\mathcal{K}_i^R$, i.e.,
\begin{equation}\label{eq:demand}
Q_{i}\left(B_{i},B_{j},p_{i},p_{j}\right)=\sum_{k\in\mathcal{K}_i^R\left(B_{i},B_{j},p_{i},p_{j}\right)}g_{k}e^{-(1+p_i)}.
\end{equation}
\end{definition}

Notice that both $\mathcal{K}_i^R$ and $Q_i$ depend on prices $(p_i,p_j)$ in Stage II and leasing decisions $(B_i,B_j)$ in Stage I.
Calculating the Realized Demands also requires considering two different pricing cases. %
%
%
\begin{enumerate}
\item \emph{Different prices} ($p_i < p_j$):
The Preferred Demands are $D_{i}(p_{i},p_{j})=Ge^{-(1+p_i)}$ and
$D_{j}(p_{i},p_{j})=0$.
%
%
\begin{itemize}

\item \emph{If Operator $i$ has enough resource} $\left(i.e., B_{i}\geq D_{i}\left(p_{i},p_{j}\right)\right)$:
all Preferred Demand will be satisfied by operator $i$. The Realized
Demands are
\begin{eqnarray*}
Q_i &=&\min(B_{i},D_{i}(p_{i},p_{j}))
=Ge^{-(1+p_i)},\\
Q_j&=&0.
\end{eqnarray*}

\item \emph{If Operator $i$ has limited resource} $\left(i.e., B_{i}<D_{i}\left(p_{i},p_{j}\right)\right)$:
since operator $i$ cannot satisfy the Preferred Demand, some demand
will be satisfied by operator $j$ if it has enough resource.
Since the realized demand
$Q_i(B_i,B_j,p_i,p_j)=B_{i}=\sum_{k\in\mathcal{K}_i^R} g_k
e^{-(1+p_i)}$, then $\sum_{k\in\mathcal{K}_i^R}g_k=B_{i}
e^{1+p_i}$. The remaining users want to purchase bandwidth from
operator $j$ with a total demand of
$\left(G-B_{i}e^{1+p_i}\right)e^{-\left(1+p_j\right)}$. Thus the
Realized Demands are
\begin{eqnarray*}
Q_i&=&\min(B_{i},D_{i}(p_{i},p_{j}))=B_{i},\\
Q_j&=&\min\left(B_{j},\left(G-B_{i}e^{1+p_i}\right)e^{-\left(1+p_j\right)}\right).
\end{eqnarray*}
\end{itemize}

\item \emph{Same prices} ($p_i = p_j=p$): both operators will attract
the same Preferred Demand $ Ge^{-(1+p)}/2$. The Realized Demands are
\begin{eqnarray*}
Q_i&=&\min\left(B_{i},D_{i}(p,p)+\max\left(D_{j}(p,p)-B_{j},0\right)\right)\\
&=&\min\left(B_{i},  \frac{G}{2e^{1+p}}+\max\left(\frac{G}{2e^{1+p}}-B_{j},0\right)\right),\\
Q_j&=&\min\left(B_{j},D_{j}(p,p)+\max\left(D_{i}(p,p)-B_{i},0\right)\right)\\
&=&\min\left(B_{j},\frac{G}{2e^{1+p}}+\max\left(\frac{G}{2e^{1+p}}-B_{i},0\right)\right).
\end{eqnarray*}
\end{enumerate}

\subsection{Operators' Pricing Competition in Stage II}
\label{Sect:stageIII} In Stage II, the two operators simultaneously
determine their prices $(p_{i},p_{j})$ considering the users'
preferred demands in Stage III, given the investment decisions
$\left(B_{i},B_{j}\right)$ in Stage I.

An operator $i$'s profit is%
\begin{equation}\label{eq:profit}
\pi_i(B_{i},B_{j},p_i,p_j)=p_i Q_i(B_{i},B_{j},p_i,p_j)-B_{i}C_{i},
\end{equation}
which is the difference between the revenue and the total cost.
Since the payment $B_{i}C_{i}$ is fixed at this stage, operator $i$
wants to maximize the revenue $p_{i}Q_{i}$.
\begin{game}[Pricing Game] \label{game:pricing}
The competition between the two operators in Stage II can be modeled
as the following game:
\begin{itemize}
\item Players: two operators $i$ and $j$.
\item Strategy space: operator $i$ can choose price $p_{i}$ from the feasible set $\mathcal{P}_i=[0,\infty)$.
Similarly for operator $j$.
\item Payoff function: operator $i$ wants to maximize the revenue $p_i Q_i(B_{i},B_{j},p_i,p_j)$.
Similarly for operator $j$.
\end{itemize}
\end{game}

At an equilibrium of the pricing game,
$(p_{i}^{\ast},p_{j}^{\ast})$, each operator maximizes its payoff
assuming that the other operator chooses the equilibrium price,
i.e.,
$$p_{i}^{\ast}=\arg\max_{p_{i}\in\mathcal{P}_{i}}p_i Q_i(B_{i},B_{j},p_i,p_j^{\ast}),\;\; i=1,2, i\neq j.$$
In other words, no operator wants to unilaterally change its pricing decision at an equilibrium.



Next we will investigate the existence and uniqueness of the pricing
equilibrium. First, we show that it is sufficient to only consider
symmetric pricing equilibrium for Game \ref{game:pricing}.

\begin{proposition}\label{thm:equalNEs}
Assume both operators lease positive bandwidth in Stage I, i.e.,
$\min\left(B_{i},B_{j}\right)>0$. If pricing equilibrium exists, it
must be symmetric\com{Then there does not exist an asymmetric
pricing equilibrium with} $p_i^*= p_j^\ast$.
\end{proposition}

The proof of Proposition \ref{thm:equalNEs} is given in Appendix
\ref{Proof_Thm1}. The intuition is that no operator will announce a
price higher than its competitor in a fear of losing its Preferred
Demand. This property significantly simplifies the search for all
possible equilibria.

\begin{figure}[tt]
\centering
\includegraphics[width=0.45\textwidth]{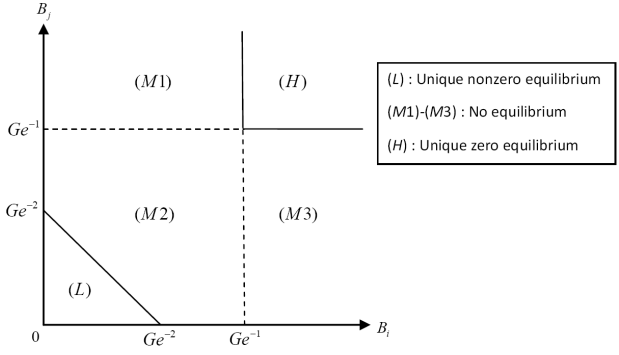}
\caption{Pricing equilibrium types in different ($B_{i},B_{j}$) regions}
\label{fig:PricingNE}
\end{figure}

Next we show that
the symmetric pricing equilibrium is a function of  ($B_i,B_j$) as shown in Fig. \ref{fig:PricingNE}.

%

\begin{theorem} \label{thm:pricingNEs}
The equilibria of the pricing game are as follows.

\begin{itemize}
\item \emph{Low Investment  Regime:} ($B_{i}+B_{j}\leq G
e^{-2}$  as in region (L) of Fig. \ref{fig:PricingNE}): there
exists a unique nonzero pricing equilibrium
\begin{equation}\label{eq:lowsupplyprice}
p_i^*(B_{i},B_{j})=p_j^*(B_{i},B_{j})=\ln\left(\frac{G}{B_{i}+B_{j}}\right)-1.
\end{equation}
The operators' profits at Stage II are
\begin{equation}\label{eq:R_IIi}\pi_{II,i}(B_{i},B_{j})=B_{i}\left(\ln\left(\frac{G}{B_{i}+B_{j}}\right)-1-C_{i}\right),
\end{equation}
\begin{equation}\label{eq:R_IIj}\pi_{II,j}(B_{i},B_{j})=B_{j}\left(\ln\left(\frac{G}{B_{i}+B_{j}}\right)-1-C_{j}\right).
\end{equation} 

\item \emph{Medium Investment Regime} ($B_{i}+B_{j}> Ge^{-2}$ and $\min(B_{i},B_{j})<Ge^{-1}$
as in regions (M1)-(M3) of Fig. \ref{fig:PricingNE}):  there is no
pricing equilibrium.

\item \emph{High Investment Regime} ($\min(B_{i}, B_{j})\geq Ge^{-1}$ as in region (H) of Fig. \ref{fig:PricingNE}):  there
exists a unique zero pricing equilibrium
\begin{equation}
p_i^*(B_{i},B_{j})=p_j^*(B_{i},B_{j})=0,
\end{equation}
and the operators' profits are negative for any positive values of $B_{i}$ and $B_{j}$.

\end{itemize}
\end{theorem}

Proof of Theorem \ref{thm:pricingNEs} is given in
Appendix~\ref{Proof_Thm2}. Intuitively, higher investments in Stage
I will lead to lower equilibrium prices in Stage II. Theorem
\ref{thm:pricingNEs} shows that the only interesting case is the low
investment regime where both operators' total investment is no
larger than $Ge^{-2}$, in which case there exists a unique positive
symmetric pricing equilibrium. Notice that same prices at
equilibrium do not imply same profits, as the operators can have
different costs ($C_i$ and $C_j$) and thus can make different
investment decisions ($B_i$ and $B_j$) as shown next.

\subsection{Operators' Leasing Strategies in Stage I}\label{subsec:stageIV}
In Stage I, the operators need to decide the leasing amounts $(B_{i},B_{j})$ to
maximize their profits. Based on Theorem \ref{thm:pricingNEs}, we
only need to consider the case where the total bandwidth of both the
operators is no larger than $G e^{-2}$.

\begin{game}[Investment Game] \label{game:investment}
The competition between the two operators in Stage I can be modeled
as the following game:
\begin{itemize}
\item Players: two operators $i$ and $j$.
\item Strategy space: the operators will choose $(B_{i},B_{j})$ from the set
$\mathcal{B}=\{(B_{i},B_j):B_i+B_j\leq G e^{-2}\}$. Notice that the
strategy space is coupled across the operators, but the operators do
not cooperate with each other.
\item Payoff function: the operators want to maximize their profits in (\ref{eq:R_IIi}) and (\ref{eq:R_IIj}), respectively.
\end{itemize}
\end{game}

At an equilibrium of the investment game, $(B_{i}^{\ast},B_{j}^{\ast})$, each operator has maximized its payoff assuming that the other operator chooses the equilibrium investment, i.e.,
\begin{equation*}
B_{i}^{\ast}=\arg\max_{0\leq B_{i}\leq Ge^{-2}-B_{j}^{\ast}}\pi_{II,i}(B_{i},B_{j}^{\ast}),\;\; i=1,2, i\neq j.
\end{equation*}

To calculate the investment equilibria of Game
\ref{game:investment}, we can first calculate operator $i$'s best
response given operator $j$'s (not necessarily equilibrium) investment decision, i.e.,
\begin{equation*}
B_{i}^{\ast}(B_{j})=\arg\max_{0\leq B_{i}\leq Ge^{-2}-B_{j}}\pi_{II,i}(B_{i},B_{j}),\;\; i=1,2, i\neq j.
\end{equation*}
%
%
By looking at operator $i$'s profit in (\ref{eq:R_IIi}), we can see
that a larger investment decision $B_i$ will lead to a smaller
price. The best choice of $B_{i}$ will achieve the best tradeoff
between a large bandwidth and a small price. \com{We can skip the
proof of the best responses (not to include that in the appendix).
However, we need to keep the proof of the equilibrium (the next
theorem). In that proof, we need to state what are the best response
functions, and prove the equilibrium starting from there. The
operators' best investment responses are shown in Appendix
\ref{Proof_Thm3} with detailed proof.}

\rev{After obtaining best investment responses of duopoly, we can
then calculate the investment equilibria, given different costs
$C_i$ and $C_j$.}

\begin{figure}[tt]
\centering
\includegraphics[width=0.4\textwidth]{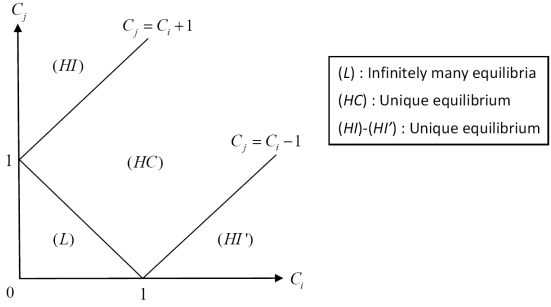}
\caption{Leasing equilibrium types in different ($C_{i},C_{j}$)
regions} \label{fig:leasingNE}
\end{figure}

\begin{theorem}\label{thm:leasingNEs}
The duopoly investment (leasing) equilibria in Stage I are
summarized as follows.
\begin{itemize}
\item \emph{Low Costs Regime} ($0<C_{i}+C_{j}\leq 1$, as region ($L$) in Fig. \ref{fig:leasingNE}):
there exists infinitely many investment equilibria characterized by
\begin{equation}\label{eq:leasing1} B_{i}^*=\rho Ge^{-2},  \textrm{  }B_{j}^*=(1-\rho)G e^{-2},
\end{equation}
where $\rho$ can be any value that satisfies
\begin{equation}\label{eq:leasing2}C_j\leq \rho\leq 1-C_i.
\end{equation}
The operators' profits are
\begin{equation}\label{eq:Lprofiti}\pi_{I,i}^L=B_{i}^*(1-C_{i}),\end{equation}
\begin{equation}\label{eq:Lprofitj}\pi_{I,j}^L=B_{j}^*(1-C_{j}),\end{equation} where  ``$L$'' denotes the
low costs regime.

\item \emph{High Comparable Costs Regime} ($C_{i}+C_{j}>1$ and $|C_{j}-C_i|\leq 1$, as region ($HC$) in Fig. \ref{fig:leasingNE}):
there exists a unique investment equilibrium
\begin{equation}\label{eq:leasingmiddlei}B_{i}^*=\frac{(1+C_{j}-C_{i})G}{2}e^{-\frac{C_{i}+C_{j}+3}{2}},\end{equation}
\begin{equation}\label{eq:leasingmiddlej}B_{j}^*=\frac{(1+C_{i}-C_{j})G}{2}e^{-\frac{C_{i}+C_{j}+3}{2}}.\end{equation}
The operators' profits are
\begin{equation}\label{eq:HCleasingi}\pi_{I,i}^{HC}=\left(\frac{1+C_{j}-C_{i}}{2}\right)^2 G e^{-\left(\frac{C_{i}+C_{j}+3}{2}\right)},\end{equation}
\begin{equation}\label{eq:HCleasingj}\pi_{I,j}^{HC}=\left(\frac{1+C_{i}-C_{j}}{2}\right)^2 G e^{-\left(\frac{C_{i}+C_{j}+3}{2}\right)},\end{equation}
where ``$HC$'' denotes the high comparable costs regime.

\item \emph{High Incomparable Costs Regime}  ($C_{j}>1+C_{i}$ or $C_{i}>1+C_{j}$, as regions ($HI$) and ($HI'$) in Fig.~\ref{fig:leasingNE}): For the case of  $C_{j}>1+C_{i}$, there exists a
unique investment equilibrium with
\begin{equation}\label{eq:leasingineqj}B_{i}^*=G
e^{-(2+C_{i})}, \textrm{ }B_{j}^*=0,\end{equation} i.e., operator
$i$ acts as the monopolist and operator $j$ is out of the market.
The operators' profits are
\begin{equation}\label{eq:HIleasing1}\pi_{I,i}^{HI}=G e^{-(2+C_{i})}, \textrm{ } \pi_{I,j}^{HI}=0,\end{equation} where  ``$HI$'' denotes the
high incomparable costs. The case of $C_{i}>1+C_{j}$ can be analyzed
similarly.
\end{itemize}
\end{theorem}

The proof of Theorem \ref{thm:leasingNEs} is given in Appendix
\ref{Proof_Thm4}.
Let us further discuss the properties of the investment equilibrium
in three different costs regimes.

\begin{figure}[tt]
\centering
\includegraphics[width=0.35\textwidth]{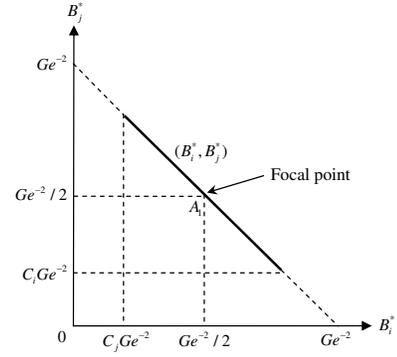}
\caption{Duopoly Leasing Focal Point with equal investments}
\label{fig:FE1}
\end{figure}

\begin{figure}[tt]
\centering
\includegraphics[width=0.35\textwidth]{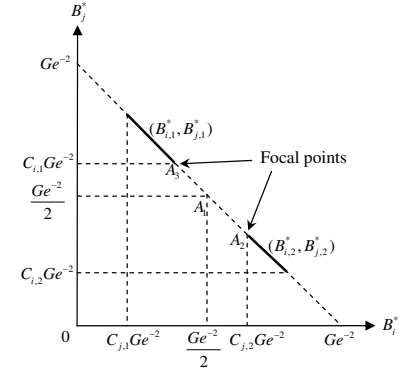}
\caption{Duopoly Leasing Focal Points with investments that have the
minimum differences}\label{fig:FE23}
\end{figure}

\begin{table*}[th]
\centering \caption{Operators' and Users' Behaviors at the Equilibria (assuming $C_i\leq C_j$)}
\begin{tabular}{|c|c|c|c|}
\hline %
\multirow{2}{*}{Costs regimes} &Low costs:   & High comparable costs:  & High incomparable costs: \\ %
&$C_{i}+C_{j}\leq 1$ & $C_{i}+C_{j}>1$ and $C_{j}-C_{i}\leq 1$& $C_{j}>1+C_{i}$\\\hline\hline %
Number of equilibria & Infinite & Unique & Unique \\ \hline %
Investment equilibria & $(\rho Ge^{-2},(1-\rho)Ge^{-2}),$ &
$\left(\frac{(1+C_{j}-C_{i})G}{2e^{\frac{C_{i}+C_{j}+3}{2}}},\frac{(1+C_{i}-C_{j})G}{2e^{\frac{C_{i}+C_{j}+3}{2}}}\right)$
& $(G e^{-(2+C_{i})},0)$ \\%
$(B_{i}^*,B_{j}^*)$ &with $C_j\leq \rho\leq (1-C_i)$ & & \\ \hline %
Pricing equilibrium ($p_i^*,p_j^*$) & $(1,1)$ & $\left(\frac{C_{i}+C_{j}+1}{2},\frac{C_{i}+C_{j}+1}{2}\right)$ &
$(1+C_{i},N/A)$ \\ \hline %
\multirow{2}{*}{Profits ($\pi_{I,i},\pi_{I,j}$)} & $\pi_{I,i}^L=\rho (1-C_i)Ge^{-2}$, & $\pi_{I,i}^{HC}=\left(\frac{1+C_{j}-C_{i}}{2}\right)^2 G e^{-\left(\frac{C_{i}+C_{j}+3}{2}\right)},$ & $\pi_{I,i}^{HI}=G e^{-(2+C_{i})},$ \\ %
& $\pi_{I,j}^L=(1-\rho)(1-C_j)Ge^{-2}$  &
$\pi_{I,j}^{HC}=\left(\frac{1+C_{i}-C_{j}}{2}\right)^2 G
e^{-\left(\frac{C_{i}+C_{j}+3}{2}\right)}$& $\pi_{I,j}^{HI}=0$
\\\hline
User $k$'s bandwidth demand & $g_ke^{-2}$ &
$g_ke^{-\left(\frac{C_i+C_j+3}{2}\right)}$ &
$g_ke^{-(2+C_i)}$\\\hline User $k$'s SNR & $e^{2}$ &
$e^{\frac{C_{i}+C_{j}+3}{2}}$ & $e^{2+C_{i}}$\\\hline User $k$'s
payoff & $g_k e^{-2}$ & $g_k
e^{-\left(\frac{C_{i}+C_{j}+3}{2}\right)}$ & $g_k
e^{-(2+C_{i})}$\\\hline
\end{tabular}
\label{tab:equilibrium}
\end{table*}

\subsubsection{Low Costs Regime ($0<C_{i}+C_{j}\leq 1$)} In this case,
both the operators have very low costs. It is the best response for
each operator to lease as much as possible. However, since the
strategy set in the Investment Game is coupled across the operators
(i.e., $\mathcal{B}=\{(B_{i},B_j):B_i+B_j\leq G e^{-2}\}$), there
exist infinitely many ways for the operators to achieve the maximum
total leasing amount $G e^{-2}$. We can further identify the focal
point, i.e., the equilibrium that the operators will agree on
without prior communications \cite{Myerson}.

For our problem, the Focal Point should be \emph{Pareto efficient}
and \emph{fair} to the operators. It is easy to check that all
investment equilibria are Pareto efficient. And fairness can be
interpreted as in terms of either equal investments or equal
profits.
Due to space limitations, we will discuss the choice of Focal Points to reach equal investments. The case of equal profits can be derived in a very similar fashion and is omitted here due to space limitations.


\com{Double check and try to put figures/tables on the same page with the corresponding main texts. You need to hid all comments before doing this final position adjustments.}

We illustrate two  types of Focal Points in Fig.~\ref{fig:FE1} and
\ref{fig:FE23}. The axes represent the equilibrium investment
amounts for two operators. The solid line segments represent the set
of infinitely many investment equilibrium. The constraints in
(\ref{eq:leasing2}) determine the starting and ending points of the
segments.

\begin{itemize}
\item Figure \ref{fig:FE1}: when $\max(C_i,C_j)\leq 1/2$, equal leasing
amount $(B_{i}^{\ast},B_j^{\ast})=(Ge^{-2}/2,Ge^{-2}/2)$ at point
$A_1$ is one of the equilibria and thus is the Focal Point.
\item Figure \ref{fig:FE23}: when $\max(C_i,C_j)>1/2$, it is not possible
for the two operators to lease the same amount at the equilibrium.
The two separate solid line segments represent the two cases of
$(C_i>1/2,C_{j}<1/2)$ and $(C_{i}<1/2,C_j>1/2)$, respectively. For
the case 1 of $(C_i>1/2,C_{j}<1/2)$ (the higher left solid line
segment), the point $A_3$ that has the smallest difference between
two equilibrium investment amounts is Focal Point, where we have
$(B_{i}^{\ast},B_j^{\ast})=\left((1-C_{i})Ge^{-2},C_{i}Ge^{-2}\right)$.
Similarly, point $A_2$ is another Focal Point for the case 2 of
$(C_{i}<1/2,C_j>1/2)$.


\end{itemize}

\subsubsection{High Comparable Costs Regime ($C_{i}+C_{j}>1$
and $|C_{j}-C_i|\leq 1$)} First, the high costs discourage the
operators from leasing aggressively, thus the total investment is less
than $G e^{-2}$. Second, the operators' costs are comparable, and thus the operator with the slightly lower cost does not have
sufficient power to drive the other operator out of the market.

\subsubsection{High Incomparable Costs Regime ($C_{j}>1+C_{i}$
or $C_{i}>1+C_{j}$)} First, the costs are high and thus the total
investment of two operators is less than $Ge^{-2}$. Second, the
costs of the two operators are so different that the operator with
the much higher cost is driven out of the market. As a result, the
remaining operator thus acts as a monopolist.


\section{Equilibrium Summary}\label{sec:Equilibrium}

Based on the discussions in Section \ref{sec:BackwardInduction}, we
summarize the equilibria
of the three-stage game in Table \ref{tab:equilibrium},
which includes the operators' investment decisions, pricing
decisions, and the resource allocation to the users. Without loss of
generality, we assume $C_i \leq C_j$ in Table \ref{tab:equilibrium}.
The equilibrium for $C_i>C_j$ can be decribed similarly.

Several interesting observations are as follows.
\begin{observation}\label{ob:linearinvestment}
 The operators' equilibrium investment decisions
$B_i^*$ and $B_j^*$ are linear in the users' aggregate
wireless
characteristics $G\left(=\sum_{k\in\mathcal{K}}g_{k}=\sum_{k\in\mathcal{K}}P_k^{max}h_k/n_0\right)$.
\end{observation}

This shows that the operators' total investment increases with the
user population, users' channel gains, and users' transmission
powers.

\begin{observation}\label{ob:indepprice}
The symmetric equilibrium price $p_i^*=p_j^*$ does not depend on
users' wireless characteristics.
\end{observation}

Observations \ref{ob:linearinvestment} and \ref{ob:indepprice} are
closely related. Since the total investment is linearly proportional
to the users' aggregate characteristics, the ``average'' equilibrium
resource allocation per user is ``constant" and does not depend on
the user population. Since resource allocation is determined by the
price, this means that the price is also independent of the user
population and wireless characteristics.

\begin{observation}\label{ob:threshold}
The operators' equilibrium investment and pricing decisions follow
simple linear threshold structures, which are easy to implement in
practice.
\end{observation}

For equilibrium investment decisions in Stage I, the feasible set of
investment costs can be divided into three regions by simple linear
thresholds as in Fig.~\ref{fig:leasingNE}. As leasing costs
increase, operators invest less aggressively; as the leasing cost
difference increases, the operator with a lower cost gradually
dominates the spectrum market. For the equilibrium pricing
decisions, the feasible set of leasing bandwidths is also divided
into three regions by simple linear thresholds as well. A meaningful
pricing equilibrium exists only when the total available bandwidth
from the two operators is no larger than a threshold (see
Fig.~\ref{fig:PricingNE}).

\begin{observation}\label{ob:demandSNR}
Each user $k$'s equilibrium demand is positive, linear in its
wireless characteristic $g_k$, and decreasing in the price. Each
user $k$ achieves the same SNR independent of $g_k$, and obtains a
payoff linear in $g_k$.
\end{observation}

Observation \ref{ob:demandSNR} shows that the users receive fair
resource allocation and QoS. Such allocation does not depend on the
wireless characteristics of the other users.

\begin{observation}\label{ob:SNRCiCj}
In the High Incomparable Costs Regime, users' \com{same} equilibrium
SNR increases with the costs $C_i$ and $C_j$, and the equilibrium
payoffs decrease with the costs.
\end{observation}

As the costs $C_i$ and $C_j$ increase, the pricing equilibrium
($p_i^\ast=p_j^\ast$) increases to compensate the loss of the
operators' profits due to increased costs. As a result, each user
will purchase less bandwidth from the operators. Since a user
spreads its total power across the entire allocated bandwidth, a
smaller bandwidth means a higher SNR but a smaller payoff.


Finally, we observe that the users achieve a high SNR at the
equilibrium. The minimum equilibrium SNR that users achieve among
the three costs regime is $e^2$. In this case, the ratio between the
high SNR approximation and Shannon capacity,
$\ln(\mathtt{SNR})/\ln(1+\mathtt{SNR})$, is larger than $94\%$. This validates our assumption on the high SNR regime. The next section, on the other hand, shows that most of the insights remain valid in the general SNR regime.


\section{Equilibrium Analysis under the General SNR Regime}\label{sec:robust}

In Sections \ref{sec:BackwardInduction} and \ref{sec:Equilibrium}, we computed the equilibria of the three-stage game based on the high SNR assumption in (\ref{eq:ratehighSNR}), and obtained five important observations (Observations 1-5). The high SNR assumption enables us to obtain closed-form solutions of the equilibria analysis and clear engineering insights.

In this section, we further consider the more general SNR regime where a user's rate is computed according to (\ref{eq:rate}) instead of (\ref{eq:ratehighSNR}). We will follow a similar
backward induction analysis, and
extend Observations \ref{ob:linearinvestment}, \ref{ob:indepprice},
\ref{ob:demandSNR}, \ref{ob:SNRCiCj}, and pricing threshold structure
of Observation \ref{ob:threshold} to the general SNR regime.

\begin{figure}[tt]
\centering
\includegraphics[width=0.35\textwidth]{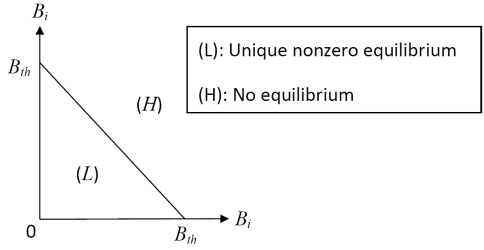}
\caption{Pricing equilibrium types in different ($B_i,B_j$) regions
for general SNR regime} \label{fig:PricingNEgeneral}
\end{figure}

We first examine the pricing equilibrium in Stage II.

\begin{theorem}\label{thm:pricinggeneralSNR}
Define $B_{th}:=0.462G$. The pricing equilibria  in the general SNR regime are as follows.
\begin{itemize}
\item \emph{Low Investment Regime} ($B_i+B_j\leq B_{th}$ as in region ($L$) of Fig.
\ref{fig:PricingNEgeneral}): there exists a unique pricing
equilibrium
\begin{multline}\label{eq:priceGeneralSNR}
p_i^*(B_i,B_j)=p_j^*(B_i,B_j)\\
=\ln\left(1+\frac{G}{B_i+B_j}\right)-\frac{G}{B_i+B_j+G}.
\end{multline}
The operators' profits at Stage II are
\begin{equation}\label{eq:iprofit_pricing}
\pi_{i}(B_i,B_j)=B_i\left[\ln\left(1+\frac{G}{B_i+B_j}\right)-\frac{G}{B_i+B_j+G}-C_i\right],
\end{equation}
\begin{equation}\label{eq:jprofit_pricing}
\pi_{j}(B_i,B_j)=B_j\left[\ln\left(1+\frac{G}{B_i+B_j}\right)-\frac{G}{B_i+B_j+G}-C_j\right].
\end{equation}
\item \emph{High Investment Regime} ($B_i+B_j>B_{th}$ as in region ($H$) of Fig.
\ref{fig:PricingNEgeneral}): there is no pricing equilibrium.
\end{itemize}
\end{theorem}

Proof of Theorem \ref{thm:pricinggeneralSNR} is given in Appendix
\ref{proofGeneralSNR}. This result is similar to Theorem
\ref{thm:pricingNEs} in the high SNR regime, and shows that the
pricing equilibrium in the general SNR regime still has a
\emph{threshold structure} in Observation \ref{ob:threshold}. Based
on Theorem \ref{thm:pricinggeneralSNR}, we are ready to prove
Observations \ref{ob:linearinvestment}, \ref{ob:indepprice},
\ref{ob:demandSNR}, and \ref{ob:SNRCiCj} in the general SNR regime.

\begin{theorem}\label{thm:ObgeneralSNR}
Observations \ref{ob:linearinvestment}, \ref{ob:indepprice},
\ref{ob:demandSNR}, and \ref{ob:SNRCiCj} in Section
\ref{sec:Equilibrium} still hold for the general SNR regime.
\end{theorem}

Proof of Theorem \ref{thm:ObgeneralSNR} is given in Appendix
\ref{proofObservations}.

\section{Impact of Operator Competition}\label{sec:Efficiency}

We are interested in understanding the impact of operator competition on the
operators' profits and the users' payoffs. As a benchmark, we will
consider the \emph{coordinated} case where both operators jointly
make the investment and pricing decisions to maximize their total
profit. In this case, there does not exists competition between the
two operators. However, it is still a Stackelberg game between a
single decision maker (representing both
operators) and the users. Then we will compare the equilibrium of
this Stackelberg game with that of the duopoly game as
in Section \ref{sec:Equilibrium}.

\subsection{Maximum Profit in the Coordinated Case}
\begin{figure}[tt]
\centering
\includegraphics[width=0.4\textwidth]{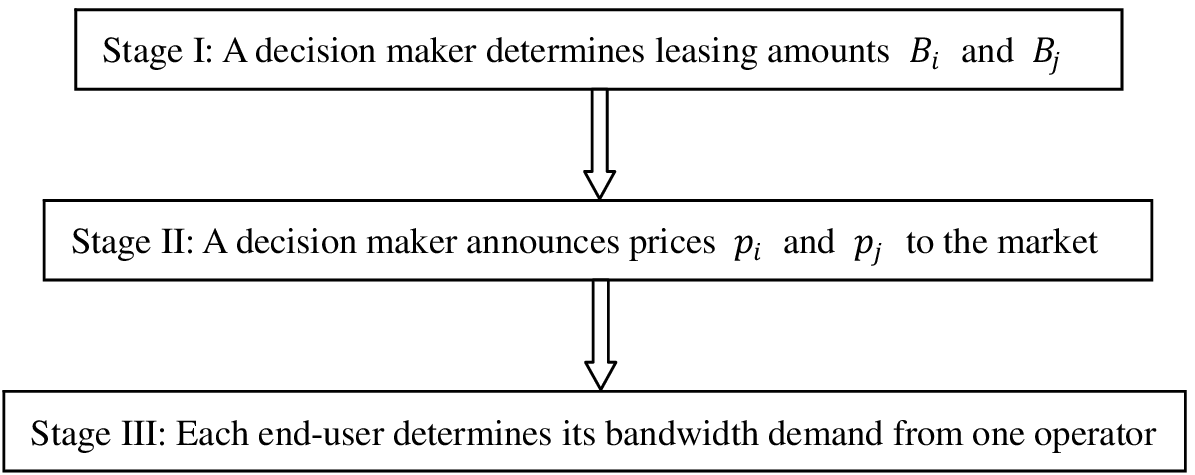}
\caption{The three-stage Stackelberg game for the coordinated
operators \com{Remove ``central'' from the first and second lines.}} \label{fig:threegame_coordinated}
\end{figure}

We analyze the coordinated case following a three stage model as
shown in Fig.~\ref{fig:threegame_coordinated}. Compared with
Fig.~\ref{fig:threestage}, the key difference here is that a single
decision maker makes the decisions in both Stages I and II. In other
words, the two operators coordinate with each other.

Again we use backward induction to analyze the three-stage game. The
analysis of Stage III in terms of the spectrum allocation among the
users is the same as in Subsection \ref{subsec:stageIII} (still
assuming the high SNR regime), and we focus on Stages II and I.
Without loss of generality, we assume that $C_i\leq C_j$.

In Stage II, the decision maker maximizes the following
total profit $T_{\pi}$ by determining prices $p_i$ and $p_j$:
\begin{equation*}
T_{\pi}(B_i,B_j,p_i,p_j)=
\pi_i(B_i,B_j,p_i,p_j)+\pi_j(B_i,B_j,p_i,p_j),%
\end{equation*}
where $\pi_i(B_i,B_j,p_i,p_j)$ is given in (\ref{eq:profit}) and
$\pi_j(B_i,B_j,p_i,p_j)$ can be obtained similarly.

%

\begin{theorem}\label{thm:Global_price}
In Stage II, the optimal pricing decisions for the coordinated
operators are as follows:
\begin{itemize}
\item If $B_i>0$ and $B_j=0$, then operator $i$ is the monopolist and announces a price
\begin{equation}\label{eq:co_price}p_i^{co}(B_i,0)=\ln\left(\frac{G}{B_i}\right)-1.\end{equation} Similar
result can be obtained if $B_i=0$ and $B_j>0$.
\item If $\min(B_i,B_j)>0$, then both operator $i$ and $j$ announce
the same price
\begin{equation}p_i^{co}(B_i,B_j)=p_j^{co}(B_i,B_j)=\ln\left(\frac{G}{B_i+B_j}\right)-1.\end{equation}

\end{itemize}
\end{theorem}

Proof of Theorem~\ref{thm:Global_price} can be found in Appendix
\ref{Proof_Thm5}. Theorem \ref{thm:Global_price} shows that both
operators will act together as a monopolist in the pricing stage.
%

Now let us consider Stage I, where the decision maker
determines the leasing amounts $B_i$ and $B_j$ to maximize the total
profit:
\begin{multline}\label{eq:Global_R_I}
\max_{B_i,B_j\geq 0}T_{\pi}(B_i,B_j)\\ =\max_{B_i,B_j\geq
0}B_i(p_i^{co}(B_i,B_j)-C_i) +B_j(p_j^{co}(B_i,B_j)-C_j),
\end{multline}
where $p_i^{co}(B_i,B_j)$ and $p_j^{co}(B_i,B_j)$ are given in
Theorem \ref{thm:Global_price}. In this case, operator $j$ will not
lease (i.e., $B_j^{co}=0$) as operator $i$ can lease with a lower
cost. Thus the optimization problem in (\ref{eq:Global_R_I})
degenerates to
$$\max_{B_i\geq 0}T_{\pi}(B_i)=\max_{B_i\geq 0}B_i(p_i^{co}(B_i,0)-C_i).$$ This leads to the following result.

\begin{theorem}\label{thm:coordinatedStageI}
In Stage I, the optimal investment decisions for the coordinated
operators are
\begin{equation}\label{eq:OptB_i}
B_i^{co}(C_i,C_j)=Ge^{-(2+C_i)}, \;\; B_j^{co}(C_i,C_j)=0,
\end{equation}
and the total profit is
\begin{equation}\label{eq:OptProfit}T_{\pi}^{co}(C_i,C_j)=G
e^{-(2+C_i)}.\end{equation}
\end{theorem}

\subsection{Impact of Competition on the Operators' Profits}

Let us compare the total profit obtained in the competitive duopoly
case (Theorem \ref{thm:leasingNEs}) and the coordinated case
(Theorem \ref{thm:coordinatedStageI}).

\subsubsection{Low Costs Regime ($0<C_i+C_j\leq 1$)} First, the
total equilibrium leasing amount in the duopoly case is
$B_i^*+B_j^*=G e^{-2}$, which is larger than the total leasing
amount $G e^{-(2+C_i)}$ in the coordinated case. In other words,
operator competition leads to a more aggressive overall investment.
Second, the total profit at the duopoly equilibria is
\begin{equation}\label{eq:Eff_AR_comp}T_{\pi}^L(C_i,C_j,\rho)=[\rho(1-C_i)+(1-\rho)(1-C_j)]G e^{-2},\end{equation}
where $\rho$ can be any real value in the set of $[C_j,1-C_i]$. Each
choice of $\rho$ corresponds to an investment equilibrium and there
are infinitely many equilibria in this case as shown in Theorem
\ref{thm:leasingNEs}. The minimum profit ratio between the duopoly
case and the coordinated case optimized over $\rho$ is
\begin{equation}\label{eq:P-Ratio^L}%
\mathtt{T_\pi-Ratio}^{L}(C_i,C_j)=\min_{\rho\in[C_j, 1-C_i]}
\frac{T_{\pi}^L(C_i,C_j,\rho)}{T_{\pi}^{co}(C_i,C_j)}.
\end{equation} Since $T_{\pi}^L(C_i,C_j,\rho)$ is increasing
in $\rho$, the minimum profit ratio is achieved  at
\begin{equation}\label{eq:Rho}\rho^*=C_j.\end{equation}
This means
\begin{equation}\label{eq:PoA_L_(Cij)}\mathtt{T_\pi-Ratio}^L(C_i,C_j)=[C_j(1-C_i)+(1-C_j)^2]e^{C_i}.\end{equation}
Although (\ref{eq:PoA_L_(Cij)}) is a non-convex function of $C_i$
and $C_j$, we can numerically compute the minimum value over all
possible values of costs in this regime
\begin{multline}\label{eq:PoA}
\min_{(C_{i},C_{j}): 0<C_{i}+C_{j}\leq
1}\mathtt{T_\pi-Ratio}^L(C_i,C_j)\\=\lim_{\epsilon\rightarrow
0}\mathtt{T_\pi-Ratio}^L(\epsilon,0.5+\epsilon)= 0.75.
\end{multline}
This means that the total profit achieved at the duopoly equilibrium
is at least 75\% of the total profit achieved in the coordinated
case under any choice of cost parameters in the Low Costs Regime.



\subsubsection{High Comparable Costs Regime ($C_i+C_j>1$ and
$C_j-C_i\leq 1$)} First, the total duopoly equilibrium leasing
amount is $B_i^*+B_j^*=G e^{-\left(\frac{C_i+C_j+3}{2}\right)}$
which is greater than $G e^{-(2+C_i)}$ of the coordinated case.
Again, competition leads to a more aggressive overall investment.
Second, the total profit of duopoly is %
\begin{equation}\label{eq:Eff_AR_comp2}%
T_{\pi}^{HC}(C_i,C_j)=\frac{1+(C_j-C_i)^2}{2}G
e^{-\frac{C_i+C_j+3}{2}}.%
\end{equation}
And the profit ratio is
\begin{multline}\label{eq:PoA_HC_Cij}%
\mathtt{T_\pi-Ratio}^{HC}(C_i,C_j)=\frac{T_{\pi}^{HC}(C_i,C_j)}{T_{\pi}^{co}(C_i,C_j)}\\=\frac{1+(C_j-C_i)^2}{2}
e^{\frac{1-(C_j-C_i)}{2}},%
\end{multline}
which is a function of the cost difference $C_j-C_i$. Let us write
it as $\mathtt{T_\pi-Ratio}^{HC}(C_j-C_i)$. We can show that it
is a convex function and achieves its minimum at
\begin{multline}
\min_{(C_{i},C_{j}): C_i+C_j>1, 0\leq C_j-C_i\leq
1}\mathtt{T_\pi-Ratio}^{HC}(C_j-C_i)\\=\mathtt{T_\pi-Ratio}^{HC}(2-\sqrt{3})=0.773.
\end{multline}

\subsubsection{High Incomparable Costs Regime ($C_j-C_i>1$)}
In this case, only one operator leases a positive amount at the
duopoly equilibrium and achieves the same profit as in the
coordinated case. The profit ratio is $1$.


We summarize the above results as follows.

\begin{theorem}[Operators' Profit Loss]
Comparing with the coordinated case, the operator competition leads to a maximum total profit loss of $25\%$ in the low costs regime.
\end{theorem}

Since low leasing costs lead to aggressive leasing decisions and
thus intensive competitions among operators,  it is not surprising
to see that the maximum profit loss happens in the low cost regime.


\subsubsection{Further Intuitions of the Low Costs Regime}

Next we explain the intuitions behind the profit ratio
$\mathtt{T_\pi-Ratio}^L(C_i,C_j)$ as in (\ref{eq:PoA_L_(Cij)}) in
the low costs regime. We can summarize the impact of costs in this
regime as two effects.
\begin{itemize}
\item \emph{Excessive Investment (EI)} effect: \rev{when the cheaper cost $C_i$
increases under a fixed cost difference $C_j-C_i$, the competition
between the operators become more intense due to the increase of
both costs $C_i$ and $C_j$. The ratio between the total leasing
amount at the duopoly equilibrium and the coordinated case tends to
increase with the costs. Such (relatively) excessive investment
leads to a higher total payment of the operators to spectrum owners
(than the coordinated case). Such effect tends to \emph{decrease}
the profit ratio with an increasing $C_{i}$. }


\item \emph{Cheaper Resource (CR)} effect: \rev{when the cheaper cost $C_i$
increases under some fixed cost difference $C_j-C_i$, the worst-case
choice of $\rho^{\ast}$ in (\ref{eq:Rho}) also increases due to the
increase of $C_j$. This leads to more investment from the spectrum
owner with cheaper cost $C_{i}$, and is closer to the decision in
the coordinated case where the operators only invest in the cheaper
resource. Such effect tends to \emph{increase} the profit ratio with
an increasing $C_{i}$. }

\end{itemize}

\begin{figure}[tt]
\centering
\includegraphics[width=0.45\textwidth]{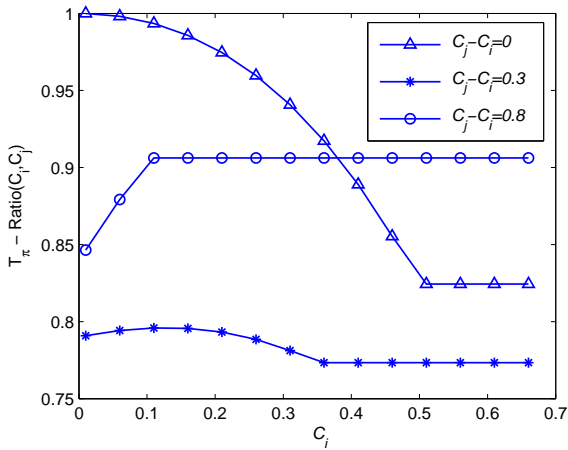}
\caption{Profit ratio $\mathtt{T_\pi-Ratio}(C_i,C_j)$ versus the
lower cost $C_i$ under different cost differences $C_j-C_i$}
\label{fig:Eff_Ci_delta}
\end{figure}

Figure \ref{fig:Eff_Ci_delta} shows three total profit ratio curves
($\mathtt{T_\pi-Ratio}(C_i,C_j)$). For each curve, the constant part
on the right hand side corresponds to the high comparable costs
($HC$) regime as in (\ref{eq:PoA_HC_Cij}), where
$\mathtt{T_\pi-Ratio}$ is a function of the cost difference
$C_{j}-C_{i}$ only. The nonlinear part on the left hand side
corresponds to the low costs regime. The interactions between the EI
and CR effects lead to different shapes of the three curves in the
low costs regime.
\begin{itemize}
\item \emph{Small cost difference} (e.g., $C_j-C_i=0$ in Fig.~\ref{fig:Eff_Ci_delta}): the Excessive Investment effect dominates. The profit ratio decreases monotonically with the cost $C_{i}$.
%

\item \emph{Medium cost difference} (e.g., $C_j-C_i=0.3$ in Fig.~\ref{fig:Eff_Ci_delta}):
both effects have comparable impacts. The profit ratio increases
first and then decreases with cost $C_{i}$.

\item \emph{Large cost difference} ($C_j-C_i=0.8$ in Fig.~\ref{fig:Eff_Ci_delta}):
the Cheaper Resource effect dominates. The profit ratio increases
monotonically with the cost $C_{i}$.


\end{itemize}

\begin{figure}[t]
\centering
\includegraphics[width=0.4\textwidth]{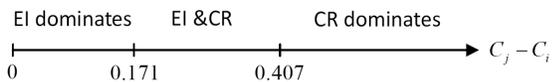}
\caption{Excessive Investment (EI) effect versus Cheaper Resource
(CR) effect under different cost difference $C_j-C_i$}
\label{fig:Twoeffect}
\end{figure}

We can numerically calculate the thresholds that separate the three
different effects interaction regions as in
Fig.~\ref{fig:Twoeffect}. Excessive Investment (EI) effect dominates
if $(C_j-C_i)\in(0,171]$, and CR effect dominates if
$(C_j-C_i)\in[0.407,1]$. The two effects have comparable impacts if
$(C_j-C_i)\in(0.171,0.407)$.

\subsection{Impact of Competition on the Users' Payoffs}


\begin{theorem}
Comparing with the coordinated case, users obtain same or higher
payoffs under the operators' competition.
\end{theorem}

By substituting (\ref{eq:OptB_i}) into (\ref{eq:co_price}), we
obtain the optimal price in the coordinated case as $1+C_i$. This
means that user $k$'s payoff equals to $g_ke^{-(2+C_i)}$ in all
three costs regimes. According to Table \ref{tab:equilibrium}, users
in the duopoly competition case have the same payoffs as in
coordinated case in the high incomparable costs regime. The payoffs
are larger in the other two costs regimes with the competitor
competition. The intuition is that operator competition in those two
regimes leads to aggressive investments, which results in lower
prices and higher user payoffs.

\section{Conclusion And Future Work}\label{sec:conclusion}
Dynamic spectrum leasing enables the secondary cognitive network
operators to quickly obtain the unused resources from the primary
spectrum owner and provide flexible services to secondary end-users. This paper studies the competition between two
cognitive operators and examines the operators' equilibrium
investment and pricing decisions as well as the users' corresponding
QoS and payoffs.

We model the economic interactions between the operators and the
users as a three-stage dynamic game. Our concrete OFDM-based  spectrum sharing
model captures the wireless heterogeneity of the users in terms of
maximum transmission power levels and channel gains. The two
operators engage in investment and pricing competitions with
asymmetric costs. We have discovered several interesting features of
the game's equilibria. For example, the duopoly's investment and
pricing decisions have nice linear threshold structures. We
also study the impact of operator competition on operators'
total profit loss and the users' payoff increases. Compared with the
coordinated case where the two operators cooperate to maximize their
total profit, we show that at the maximum profit loss due to competition is no larger than $25\%$. We also show that the users always
benefit from competition by achieving the same or better payoffs. Although we have focused on the high SNR regime when obtaining closed-form solutions, we show that most engineering insights summarized in Section \ref{sec:Equilibrium} still hold in the general SNR regime.

There are several possible ways to extend the results here. We can
consider the case where the operators can also obtain resource
through spectrum sensing as in \cite{report_Infocom}. Compared with
leasing, sensing is cheaper but the amount of useful spectrum is
less predictable due to the primary users' stochastic traffic. With
the possibility of sensing, we need to consider a four-stage dynamic
game model. We can also consider the case where users might
experience different channel conditions when they choose different
providers, e.g., when they need to communicate with the base
stations of the operators. Competition under such channel
heterogeneity has been partially considered in
\cite{GajicHuangRimoldi2009} without considering the cost of
spectrum acquisition.


\appendix
\subsection{Proof of Proposition \ref{thm:equalNEs}}\label{Proof_Thm1}

\begin{figure}[tt]
\centering
\includegraphics[width=0.4\textwidth]{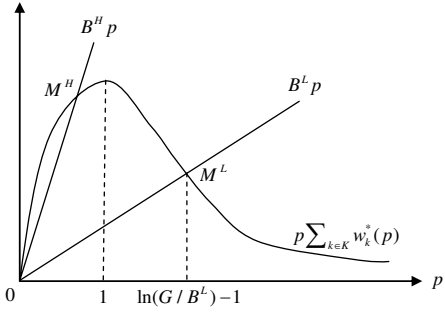}
\caption{Monopolist's revenue given low or high supply in pricing
stage} \label{fig:monopolist}
\end{figure}


If the two operators announce different prices, then the operator
with the lower price attracts all the users' demand and essentially
acts as a monopolist. We will first summarize the pricing behavior
of a monopolist, and detailed derivations are given in
\cite{report_Infocom}. After that, we will show the main proof.

\subsubsection{Monopolist's optimal pricing strategy}
Given a fixed leasing amount $B$, the monopolist wants to choose the
price $p$ to maximize its revenue.
Denote the demand of user $k$ as $w_k^*(p)$, and thus the total
demand is ${\sum_{k\in\mathcal{K}}w_k^*(p)}$. The revenue is
$p\min\left(B,{\sum_{k\in\mathcal{K}}w_k^*(p)}\right)$.
In Fig.~\ref{fig:monopolist}, the nonlinear curve represents the
function $p \sum_{k\in\mathcal{K}}w_k^*(p)$. The other two linear
curves represent two representative values of $p B$.
%
%
%
To maximize the revenue, we will have the following two cases:
\begin{itemize}
\item \emph{Monopolist's low supply regime:} if $B\leq G e^{-2}$ (e.g., $B^L$ in Fig.~\ref{fig:monopolist}),
then it is optimal to choose a price such that supply equals to
demand,
$$p^*(B)=\ln\left(\frac{G}{B}\right)-1.$$
\item \emph{Monopolist's high supply regime:} if $B\geq G e^{-2}$ (e.g., $B^{H}$ in Fig.~\ref{fig:monopolist}),
then it is optimal to choose a price such that supply exceeds
demand,
$$p^*(B)=1.$$
\end{itemize}

\begin{figure}
\centering
\includegraphics[width=0.3\textwidth]{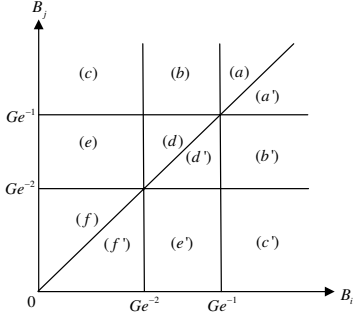}
\caption{Different ($B_{i},B_{j}$) regions} \label{fig:Blij_Proof}
\end{figure}

\subsubsection{Main proof} Now let us consider the two operator case. Suppose that there exists an equilibrium ($p_i^*,p_j^*$) where
$p_i^\ast\neq p_j^\ast$. Without loss of generality, we assume that
$B_{i}\leq B_{j}$. In the following analysis, we examine all
possible ($B_{i},B_{j}$) regions labeled (\emph{a})-(\emph{f}) as
shown in Fig. \ref{fig:Blij_Proof}.

\begin{itemize}
\item[(a)] If $B_{j}\geq B_{i} \geq G e^{-1}$, then both the operators have adequate bandwidths to
cover the total preferred demand. This is because the total
preferred demand to an operator $i$ has the maximum value of
$D_{i}(0,p_{j})=Ge^{-1}$. Thus any operator announcing a lower price
will attract all the demand. The operator charging a higher price
has no realized demand , and thus has the incentive to reduce the
price until no larger than other price. Thus unequal price is not an
equilibrium.

\item[(b)] If $G e^{-2}<B_{i}<Ge^{-1}\leq B_{j}$, operator $i$ will not announce a price higher than
operator $j$ for the same reason as in case ($a$). Furthermore, the
operator $j$ will not announce a price $p_j>1$. Otherwise, the
operator $i$ will act like a monopolist by setting $p_i=1$ to
maximize its revenue and leave no realized demand to operator $j$.
Thus we conclude that $p_i^*<p_j^*\leq 1$. But operator $i$wants to
set price $p^\ast_i = p^\ast_j -\epsilon$ where $\epsilon>0$ is
infinitely small, and thus can not reach an equilibrium.

\item[(c)] If $B_{i}\leq G e^{-2}<G e^{-1}\leq B_{j}$, then operator $i$ will
not announce a price higher than operator $j$ as in case ($a$). Also
operator $j$ will not charge a price
$p_j^*>\ln\left(\frac{G}{B_{i}}\right)-1$, otherwise operator $i$
will act like a monopolist by setting
$p_i=\ln\left(\frac{G}{B_{i}}\right)-1$ to maximize its revenue and
leave no realized demand to operator $j$. Thus we conclude that
$p_i^*<p_j^*\leq \ln\left(\frac{G}{B_{i}}\right)-1$. However, the
operator $i$ wants to set price $p^\ast_i = p^\ast_j -\epsilon$
where $\epsilon>0$ is infinitely small, and thus can not reach an
equilibrium.

\item[(d)] If $G e^{-2}\leq B_{i}\leq B_{j}< Ge^{-1}$, duopoly will not announce price $\max(p_i^*,p_j^*)>1$.
Thus we have either $p_i^*<p_j^*\leq 1$ or $p^\ast_j <p^\ast_i \leq
1$. In both cases, the operator with the higher price wants to
reduce the price to be just a little bit smaller than the other
operator's, and thus can not reach an equilibrium.
%

\item[(e)] If $B_{i}\leq G e^{-2}\leq B_{j}<Ge^{-1}$, then we have $p_i^*\leq 1$ and $p_j^*\leq
\ln\left(\frac{G}{B_{i}}\right)-1$. Thus we have either
$p_i^*<p_j^*\leq \ln\left(\frac{G}{B_{i}}\right)-1$ or
$p_j^*<p_i^*\leq1$. Similar as (d), an equilibrium can not be
reached.

\item[(f)] If $B_{i}\leq B_{j}\leq Ge^{-2}$, then we have $p_i^*\leq
\ln\left(\frac{G}{B_{j}}\right)-1$ and $p_j^*\leq
\ln\left(\frac{G}{B_{i}}\right)-1$. Thus we have either
$p_i^*<p_j^*\leq \ln\left(\frac{G}{B_{i}}\right)-1$ or
$\ln\left(\frac{G}{B_{j}}\right)-1\geq p_i^*>p_j^*$. In both cases,
the operator with the higher price wants to reduce the price to be
just a little bit smaller than the other operator's, and thus can
not reach an equilibrium.
\end{itemize}

Similar analysis can be extended to  regions ($a'$)-($f'$) in
Fig.~\ref{fig:Blij_Proof}. Thus in all all possible ($B_{i},B_{j}$)
regions, there doesn't exist a pricing equilibrium such that
$p_i^*\neq p_j^*$. \hfill$\rule{2mm}{2mm}$

\subsection{Proof of Theorem \ref{thm:pricingNEs}}\label{Proof_Thm2}

Assume, without loss of generality, that $B_{i}\leq B_{j}$. Based on
Proposition \ref{thm:equalNEs}, in the following analysis we examine
all possible ($B_{i},B_{j}$) regions labeled (a)-(f) in Fig.
\ref{fig:Blij_Proof}, and check if there exists a symmetric pricing
equilibrium (i.e., $p_i^*=p_j^*$) in each region.

\begin{itemize}
\item[(a)] If $B_{j}\geq B_{i} \geq G e^{-1}$, both the operators have adequate bandwidths to
cover the total preferred demand which reaches its maximum $G
e^{-1}$ at zero price.

\begin{itemize}
\item if $p_i^*=p_j^*>0$, then operator $i$ attracts and realizes half of the total preferred demand.
But when operator $i$ slightly decreases its price, it attracts and
realizes the total preferred demand,
and thus doubles its revenue.

\item if $p_i^*=p_j^*=0$, any operator can not attract or realize any preferred demand by
unilaterally deviating from (increasing) its price.
\end{itemize}
Hence, $p_i^*=p_j^*=0$ is the unique equilibrium in region ($a$).

\item[(b-c)] If $B_{i}\leq G e^{-2}<Ge^{-1}\leq B_{j}$ or $G e^{-2}<B_{i}<Ge^{-1}\leq B_{j}$, operator $j$ has adequate bandwidth while operator $i$ only
has limited bandwidth.

\begin{itemize}
\item if $p_i^*=p_j^*>0$, then operator $j$ will slightly reduce its price to attract and realize
the total preferred demand.

\item if $p_i^*=p_j^*=0$, then operator $j$ will increase its price
and still have positive realized demand. This is because operator
$i$ does not have enough supply to satisfy the total preferred
demand.
\end{itemize}
Hence, there doesn't exist an equilibrium in regions (\emph{b-c}).


\item[(d-e)] If $Ge^{-2}\leq B_{i}\leq B_j< Ge^{-1}$ or $B_{i}\leq G e^{-2}\leq B_{j}<Ge^{-1}$, we have shown in the proof of
Proposition \ref{thm:equalNEs} that possible pricing equilibrium
will not exceed 1. We find possible pricing equilibrium given
operator $j$'s leasing amount.

\begin{itemize}
\item if $p_i^*=p_j^*>
\ln\left(\frac{G}{B_{j}}\right)-1$, then operator $j$ has enough
bandwidth to cover the total preferred demand and it will slightly
decrease its price to attract a larger preferred demand.

\item if $p_i^*=p_j^*\leq
\ln\left(\frac{G}{B_{j}}\right)-1$, then operator $j$ has limited
bandwidth and it will make decision depending on operator $i$'s
supply.

\begin{itemize}
\item if $B_{i}\leq G e^{-(1+p_j^*)}/2$,
then operator $j$ will slightly decrease its price if $B_{i}+B_{j}>G
e^{-(1+p_j^*)}$, or increase its price to $1$ if $B_{i}+B_{j}\leq G
e^{-(1+p_j^*)}$.

\item if $B_{i}> G e^{-(1+p_j^*)}/2$,
then operator $j$ will slightly reduce its price.
\end{itemize}
\end{itemize}
Hence, there doesn't exist a pricing equilibrium in regions
(\emph{d-e}).


\item[(f)] If $B_{i}\leq B_{j}\leq Ge^{-2}$, we will first show that total supply
equals total preferred demand at any possible equilibrium (i.e.,
$p_i^*=p_j^*=\ln\left(\frac{G}{B_{i}+B_{j}}\right)-1$).

\begin{itemize}
\item Suppose that at an equilibrium $p_i^*=p_j^*<\ln\left(\frac{G}{B_{i}+B_{j}}\right)-1$ and thus the total supply
is \emph{less} than the total preferred demand. Then operator $j$
will slightly increase its price without changing much its realized
demand, and thus receive a greater revenue.

\item Suppose that at an equilibrium $p_i^*=p_j^*\geq\ln\left(\frac{G}{B_{i}+B_{j}}\right)-1$ and thus the total supply
is \emph{greater} than the total preferred demand. Thus we have
$B_{j}>Ge^{-(1+p_j^*)}/2$. Operator $j$ will slightly reduce its
price to attract much more preferred demand and receive a greater
revenue.
\end{itemize}
Thus we have $p_i^*=p_j^*=\ln\left(\frac{G}{B_{i}+B_{j}}\right)-1$
at any possible equilibrium. Then we check if such ($p_i^*,p_j^*$)
is an equilibrium for the following two cases.
\begin{itemize}
\item If $B_{i}+B_{j}>Ge^{-2}$, then we have $p_i^*=p_j^*<1$. Since operator $j$ already has its individual supply
equal to its realized demand, then operator $i$ acts as a monopolist
serving its own users in the monopolist's high investment regime in
the proof of Proposition \ref{thm:equalNEs}. Then operator $i$ will
increase its price to 1.
\item If $B_{i}+B_{j}\leq Ge^{-2}$, then
we have $p_i^*=p_j^*\geq1$. Each operator acts as a monopolist
serving its own users in the monopolist's low investment regime in
the proof of Proposition \ref{thm:equalNEs}. And it's optimal for
each operator to stick with its current price.
\end{itemize}
Thus there exists a unique pricing equilibrium
$p_i^*=p_j^*=\ln\left(\frac{G}{B_{i}+B_{j}}\right)-1$ for the low
investment regime $B_{i}+B_{j}\leq{G}e^{-2}$ in region ($f$).
\end{itemize}
The same results can be extended to symmetric regions
($a'$)-($f'$) in Fig.~\ref{fig:Blij_Proof}.
\hfill$\rule{2mm}{2mm}$

\begin{table*}
\centering \caption{Best Investment Response $B_{i}^*({B_{j}})$ of
operator $i$ in Stage I}
\begin{tabular}{|p{2.5in}|p{1.6in}|p{1.6in}|}
\hline \multicolumn{1}{|c|}{Response $B_{i}^*({B_{j}})$}
&\multicolumn{1}{|c|}{Low individual cost $0<C_{i}\leq 1$}
&\multicolumn{1}{|c|}{High individual cost $C_{i}> 1$}
\\\hline Small competitor's decision $B_{j}<C_{i}G e^{-2}$
& the solution to $\frac{\partial \pi_{II,i}(B_{i},B_{j})}{\partial
B_{i}}=0$ & N/A
\\\hline Large competitor's decision $B_{j}\geq C_{i}G e^{-2}$ &$G
e^{-2}-B_{j}$& N/A
\\\hline Small competitor's decision $B_{j}<G e^{-(1+C_{i})}$ &
N/A & the solution to $\frac{\partial
\pi_{II,i}(B_{i},B_{j})}{\partial B_{i}}=0$\\\hline Large
competitor's decision $B_{j}\geq G e^{-(1+C_{i})}$ & N/A &
$0$\\\hline
\end{tabular}
\tabcolsep 5mm \label{tab:leasingresponse}
\end{table*}

\subsection{The Operators' Best Investment Responses with Proof}\label{Proof_response}

\com{We can delete this proof.}

Due to the concavity of profit $\pi_{II,i}(B_{i},B_{j})$ in $B_{i}$,
we can obtain the best response function (i.e., best choice of
$B_{i}$ given fixed $B_{j}$) by checking the first order condition.
The best response of operator $i$ depends on cost $C_i$ and the
leasing decision of its competitor, $B_j$.
Operator $i$'s best response investment is summarized in Table
\ref{tab:leasingresponse}.
Operator $j$'s best response can be calculated similarly.

\emph{Proof.} Since $\pi_{II,i}(B_{i},B_{j})$ in (\ref{eq:R_IIi}) is
a concave function of $B_{i}$, it is enough to check the first order
condition as well as the boundary condition. We have

$$\frac{\partial \pi_{II,i}(B_{i},B_{j})}{\partial B_{i}}=\ln\left(\frac{
G}{B_{i}+B_{j}}\right)-\frac{B_{i}}{B_{i}+B_{j}}-1-C_{i}.$$ Its
values at the boundary of operator $i$'s strategy space are
$$\frac{\partial \pi_{II,i}(B_{i},B_{j})}{\partial
B_{i}}\mid_{B_{i}=0}=\ln\left(\frac{G}{B_{j}}\right)-1-C_{i},
$$ and $$\frac{\partial \pi_{II,i}(B_{i},B_{j})}{\partial
B_{i}}\mid_{B_{i}=Ge^{-2}-B_{j}}=\frac{B_{j}}{G e^{-2}}-C_{i},$$
both of which are dependent on its competitor $j$'s strategy $B_{j}$
and the cost $C_{i}$. Thus we derive operator $i$'s best response
for different costs $C_{i}$ and operator $j$'s strategies as
follows.

\begin{itemize}
\item \emph{Low individual cost regime} ($C_{i}\leq 1$), then
$$\frac{\partial \pi_{II,i}(B_{i},B_{j})}{\partial
B_{i}}\mid_{B_{i}=0}\geq 0,$$ i.e., operator $i$ is encouraged to
lease positive amount. This is because that in low investment regime
the pricing equilibrium in (\ref{eq:lowsupplyprice}) is always
larger than 1 and thus larger than $C_i$, and the profit per unit
leased bandwidth is positive.
\begin{itemize}
\item \emph{Large Competitor's Decision} ($B_{j}\geq C_{i}{Ge^{-2}}$), then $$\frac{\partial \pi_{II,i}(B_{i},B_{j})}{\partial
B_{i}}\mid_{B_{i}=G e^{-2}-B_{j}}\geq 0,$$ i.e., the large leasing
amount of operator $j$ already makes the pricing equilibrium in
(\ref{eq:lowsupplyprice}) very low (but still larger than 1). And
operator $i$'s best response is to lease as much bandwidth as
possible
$$B_{i}^*(B_{j})=G e^{-2}-B_{j},$$ which will only leads to a relatively small decrease of price.
\item \emph{Small Competitor's Decision} ($B_{j}< C_{i}{Ge^{-2}}$), then $$\frac{\partial \pi_{II,i}(B_{i},B_{j})}{\partial
B_{i}}\mid_{B_{i}=G e^{-2}-B_{j}}<0,$$ i.e., operator $i$ will not
lease aggressively to avoid making the price too low. Its best
response $B_{i}^*(B_{j})$ is the unique solution to $$\frac{\partial
\pi_{II,i}(B_{i},B_{j})}{\partial B_{i}}=0, $$ which lies in the
strict interior of $[0,G e^{-2}-B_{j})$.
\end{itemize}

\item \emph{High individual cost regime} ($C_{i}>1$), then
$$\frac{\partial \pi_{II,i}(B_{i},B_{j})}{\partial
B_{i}}\mid_{B_{i}=G e^{-2}-B_{j}}<0,$$ i.e., the high cost makes
operator $i$ not lease the maximum possible value.
\begin{itemize}
\item If \emph{Large Competitor's Decision} ($B_{j}\geq G e^{-(1+C_i)}$), then $$\frac{\partial \pi_{II,i}(B_{i},B_{j})}{\partial
B_{i}}\mid_{B_{i}=0}\leq 0,$$ i.e., the competitor $j$'s large
leasing amount makes the price low. Together with the high leasing
cost, it is optimal for operator $i$ not to lease anything. Thus we
have
$$B_i^*(B_j)=0.$$
\item \emph{Small Competitor's Decision} ($B_{j}<G e^{-(1+C_{i})}$), then $$\frac{\partial \pi_{II,i}(B_{i},B_{j})}{\partial
B_{i}}\mid_{B_{i}=0}>0,$$ i.e., operator $j$'s limited leasing
amount enables operator $i$ to lease positive amount despite of the
large leasing cost. And operator $i$'s best response
$B_{i}^*(B_{j})$ is the unique solution to
$$\frac{\partial\pi_{II,i}(B_i,B_j)}{\partial B_i}=0,$$ which lies in the strict interior of $
(0,G e^{-2}-B_{j})$.
\end{itemize}
\end{itemize}
\hfill$\rule{2mm}{2mm}$
%

\subsection{Proof of Theorem \ref{thm:leasingNEs}}\label{Proof_Thm4}

\com{We need to keep this proof and add the expressions of the best
response inside. } The best investment response of operator $i$ is
summarized in Table \ref{tab:leasingresponse} with detailed proof in
Appendix \ref{Proof_response}.
\begin{figure}[tt]
\centering
\includegraphics[width=0.25\textwidth]{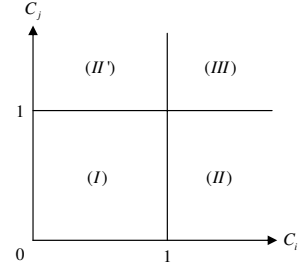}
\caption{Different ($C_{i},C_{j}$) regions}
\label{fig:leasingNE_Proof}
\end{figure}
An investment equilibrium ($B_i^*,B_j^*$) corresponds to a fixed
iteration point of two functions $B_i^*(B_j)$ and $B_j^*(B_i)$.
In the following analysis, we examine all possible costs ($C_i,C_j$)
regions labeled ($I$)-($III$) in Fig.~\ref{fig:leasingNE_Proof}, and
check if there exists any equilibrium in each region.

\begin{itemize}
\item[(I)] If $C_{i}\leq 1$ and $C_{j}\leq 1$, both the operators are in low individual cost regime.
\begin{itemize}

\item If $B_{i}^*\geq C_{j}G e^{-2}$ and $B_{j}^*\geq C_{i}Ge^{-2}$, there exist
infinitely many investment equilibria characterized by
(\ref{eq:leasing1}) and (\ref{eq:leasing2}). Since $B_i^*\geq
C_jGe^{-2}$ and $B_j^*\geq C_iGe^{-2}$, $C_i+C_j\leq 1$ is further
required for existence of equilibria.

\item If $B_{i}^*< C_{j}G e^{-2}$ and $B_{j}^*\geq C_{i}G
e^{-2}$, then by solving equations $B_{i}^*(B_j^*)=Ge^{-2}-B_j^*,$
and
$$\frac{\partial \pi_{II,j}(B_{i},B_{j})}{\partial
B_{j}}\mid_{B_{i}=B_{i}^*,B_{j}=B_{j}^*}=0,$$ we have
$B_{i}^*=C_{j}Ge^{-2}$ and $B_{j}^*=(1-C_{j})Ge^{-2}$. But the value
of $B_{i}^*$ is not smaller than $C_{j}G e^{-2}$.

\item If $B_{i}^*\geq C_{j}Ge^{-2}$ and $B_{j}^*< C_{i}Ge^{-2}$, we can also show that there does not
exist any equilibrium in this case by a similar argument as above.

\item If $B_{i}^*<C_{j}Ge^{-2}$ and $B_{j}^*< C_{i}Ge^{-2}$, then by solving equations $$\frac{\partial
\pi_{II, i}(B_{i},B_{j})}{\partial
B_{i}}\mid_{B_{i}=B_{i}^*,B_{j}=B_{j}^*}=0,$$
$$\frac{\partial \pi_{II, j}(B_{i},B_{j})}{\partial
B_{j}}\mid_{B_{i}=B_{i}^*,B_{j}=B_{j}^*}=0,$$ we have $B_{i}^*$ in
(\ref{eq:leasingmiddlei}) and $B_{j}^*$ in
(\ref{eq:leasingmiddlej}). And $C_i+C_j>1$ is further required for
existence of this equilibrium.
\end{itemize}
Hence, in region ($I$), there exist infinitely many equilibria
satisfying (\ref{eq:leasing1}) and (\ref{eq:leasing2}) when
$C_i+C_j\leq 1$, and there exists a unique equilibrium satisfying
(\ref{eq:leasingmiddlei}) and (\ref{eq:leasingmiddlej}) when
$C_i+C_j>1$.

\item[(II)] If $C_{i}>1$ and $0<C_j\leq 1$, operator $i$ is in high individual cost regime and operator $j$
is in low individual cost regime.
\begin{itemize}
\item If $B_{i}^*\geq C_{j}Ge^{-2}$ and $B_{j}^*\geq Ge^{-(1+C_i)}$, then we have
$B_{i}^*=0$ and $B_{j}^*=G e^{-2}$. But the value of $B_i^*$ is not
greater than $C_{j}Ge^{-2}$.

\item If $B_{i}^*\geq C_{j}Ge^{-2}$ and $B_{j}^*<Ge^{-(1+C_i)}$, then by solving
equations $B_j^*(B_i^*)=Ge^{-2}-B_i^*,$ and $$\frac{\partial\pi_{II,
i}(B_{i},B_{j})}{\partial
B_{i}}\mid_{B_{i}=B_{i}^*,B_{j}=B_{j}^*}=0,$$ we have
$B_i^*=(1-C_i)Ge^{-2}$ and $B_j^*=C_iGe^{-2}$. But the value of
$B_j^*$ is not less than $Ge^{-(1+C_i)}$.

\item If $B_{i}^*< C_{j}G e^{-2}$ and $B_{j}^*\geq Ge^{-(1+C_i)}$, then by solving equations
$B_i^*(B_j^*)=0,$ and $$\frac{\partial \pi_{II,
j}(B_{i},B_{j})}{\partial
B_{j}}\mid_{B_{i}=B_{i}^*,B_{j}=B_{j}^*}=0,$$ we have $B_{i}^*=0$
and $B_{j}^*=Ge^{-(2+C_j)}$. And $C_i>1+C_j$ is further required for
existence of this equilibrium.

\item $B_{i}^*< C_{j}Ge^{-2}$ and $B_{j}^*< Ge^{-(1+C_i)}$, then by
solving equations $$\frac{\partial \pi_{II,
i}(B_{i},B_{j})}{\partial
B_{i}}\mid_{B_{i}=B_{i}^*,B_{j}=B_{j}^*}=0,$$
$$\frac{\partial \pi_{II, j}(B_{i},B_{j})}{\partial
B_{j}}\mid_{B_{i}=B_{i}^*,B_{j}=B_{j}^*}=0,$$ we have $B_{i}^*$ in
(\ref{eq:leasingmiddlei}) and $B_{j}^*$ in
(\ref{eq:leasingmiddlej}). And $C_i\leq 1+C_j$ is further required
for existence of this equilibrium.

\end{itemize}
Hence, in region (II), there exists a unique investment equilibrium
($B_{i}^*,B_{j}^*$) satisfying (\ref{eq:leasingmiddlei}) and
(\ref{eq:leasingmiddlej}) when $C_i\leq 1+C_j$, and there exists a
unique equilibrium satisfying $B_i^*=0$ and $B_j^*=Ge^{-(2+C_j)}$
when $C_i>1+C_j$.

\item[(III)] If $C_{i}>1$ and $C_{j}>1$, then both the operators are in high individual cost
regime.
\begin{itemize}
\item If $B_{i}^*<Ge^{-(1+C_{j})}$ and $B_{j}^*<G e^{-(1+C_{i})}$,
then by solving equations $$\frac{\partial
\pi_{II,i}(B_{i},B_{j})}{\partial
B_{i}}\mid_{B_{i}=B_{i}^*,B_{j}=B_{j}^*}=0,$$
$$\frac{\partial \pi_{II,j}(B_{i},B_{j})}{\partial
B_{j}}\mid_{B_{i}=B_{i}^*,B_{j}=B_{j}^*}=0,$$ we have $B_{i}^*$ in
(\ref{eq:leasingmiddlei}) and $B_{j}^*$ in
(\ref{eq:leasingmiddlej}). And $C_{i}-1<C_{j}<C_{i}+1$ is further
required for existence of this equilibrium.

\item If $B_{i}^*<Ge^{-(1+C_{j})}$ and $B_{j}^*\geq Ge^{-(1+C_{i})}$, then by solving
equations $B_{i}^*(B_j^*)=0,$ and $$\frac{\partial
\pi_{j}(B_{i},B_{j})}{\partial
B_{j}}\mid_{B_{i}=B_{i}^*,B_{j}=B_{j}^*}=0,$$ we have $B_i^*=0$ and
$B_{j}^*=Ge^{-(2+C_j)}$. And $C_{j}\leq C_{i}-1$ is further required
for existence of this equilibrium.

\item If $B_{i}^*\geq G e^{-(1+C_{j})}$ and $B_{j}^*<Ge^{-(1+C_{i})}$, then we can similarly show that
there exists a unique equilibrium $B_i^*=Ge^{-(2+C_i)}$ and
$B_j^*=0$ only when $C_{j}\geq C_{i}+1.$

\item If $B_{i}^*\geq Ge^{-(1+C_{j})}$ and $B_{j}^*\geq Ge^{-(1+C_{i})}$, then we have $B_{i}^*=0$ and
$B_{j}^*=0$. However, the value of $B_i^*$ is not greater than
$Ge^{-(1+C_j)}$.
\end{itemize}
Hence, in region (III), there exists a unique equilibrium satisfying
(\ref{eq:leasingmiddlei}) and (\ref{eq:leasingmiddlej}) when
$C_{i}-1<C_{j}<C_{i}+1$; there exists a unique equilibrium
satisfying $B_{i}^*=0$ and $B_{j}^*=Ge^{-(2+C_j)}$ when $C_{j}\leq
C_{i}-1$; and there exists a unique equilibrium with
$B_{i}^*=Ge^{-(2+C_i)}$ and $B_{j}^*=0$ when $C_{j}\geq C_{i}+1$.

The same results can be extended to symmetric region ($II'$) in Fig.
\ref{fig:leasingNE_Proof}.
\end{itemize}
\hfill$\rule{2mm}{2mm}$

\subsection{Proof of Theorem \ref{thm:pricinggeneralSNR}}\label{proofGeneralSNR}

In the following analysis, we will first derive the users' optimal
behaviors in general SNR regime under a single operator case
(monopoly), and then summarize the monopolist's optimal pricing
decision. After that, we prove the symmetric pricing structure for
duopoly, and find the pricing equilibrium.

\subsubsection{The users' optimal behaviors in general SNR regime}\label{subsec:users}
Let us write the price announced by the monopolist by $p$, and the
investment amount by $B$. By demanding bandwidth $w_k$, a user $k$'s
payoff function in the general SNR regime is
\begin{equation}\label{eq:userpayoff_general}
u_k(p,w_k)=w_k \ln\left(1+\frac{g_k}{w_k}\right)-p w_k.
\end{equation}
The optimal demand $w_k^*(p)$ that maximizes
(\ref{eq:userpayoff_general}) is
\begin{equation}\label{eq:userdemand_generalSNR}w_k^*(p)=g_k/H(p),\end{equation} where $H(p)$ is the unique
positive solution to $F(p,Q):=\ln(1+H)-\frac{H}{1+H}-p=0$. The
inverse function of $H(p)$ is $p(H)=\ln(1+H)-\frac{H}{1+H}$. By
applying the implicit function theorem, we can obtain the first
derivative of function $H(p)$ over $p$ as
\begin{equation}
H'(p)=-\frac{\partial F(p,H)/\partial p}{\partial F(p,H)/\partial
p}=\frac{(1+H(p))^2}{H(p)},
\end{equation}
which is always positive. Hence, $H(p)$ is increasing in $p$.

User $k$'s optimal payoff is
\begin{equation}\label{eq:payoff_generalSNR}
u_k(p,w_k^*(p))=\frac{g_k}{H(p)}[\ln(1+H(p))-p].
\end{equation}
As a result, user $k$'s optimal SNR equals $g_k/w_k^*(p)=H(p)$ and
is \emph{user-independent}.

\subsubsection{Monopolist's optimal pricing strategy}
The users' total preferred demand equals $G/H(p)$, and the
operator's pricing problem in Stage II is to maximize its revenue
$R(B,p)=p \min(B,G/H(p))$ by optimally choosing a price. Let us
define $S(p)=p B$ and $D(p)=G p/H(p)$.

The first derivative of $D(p)$ over $p$ is
$$D'(p)=\frac{G[2H^2(p)+H(p)-(1+H(p))^2
\ln(1+H(p))]}{H^3(p)},$$ which is positive when $0\leq p\leq 0.468$
and is negative when $p>0.468$. Notice that $D'(p)$ approaches to
positive infinity when $p$ goes to 0.

The second derivative of $D(p)$ over $p$ can be shown to
be negative when $0\leq p\leq 1.266$ and positive when $p>1.266$.
Thus $D(p)$ is concave in its increasing part with $0\leq p\leq
0.468$.

\begin{figure}[tt]
\centering
\includegraphics[width=0.3\textwidth]{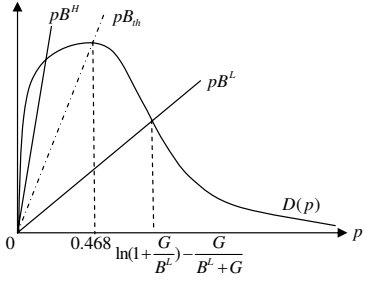}
\caption{Different relations between $D(p)$ and $S(p)$ in general
SNR regime} \label{fig:DintersectsS}
\end{figure}

Fig. \ref{fig:DintersectsS} illustrates the all possible relation
between $D(p)$ and $S(p)$ which is a linear function of $p$. The
nonlinear curve represents the function $D(p)$. The two linear solid
lines represent two representative values of $p B$. The other linear
dashed line is with threshold slope $B_{th}:=0.462G$ that intersects
$D(p)$ at its maximum value. Note that $S(p)$ always intersects
$D(p)$ since we have shown that the slope of $D(p)$ at $p=0$ becomes
positive infinity. To maximize the revenue, we will have the
following two pricing cases:
\begin{itemize}
\item \emph{Monopolist's low investment regime:} if $B\leq B_{th}$ (e.g., $B^L$ in Fig.
\ref{fig:DintersectsS}), then it is optimal to choose a price such
that the supply equals demand (i.e., $B=G/H(p)$),
\begin{equation}\label{eq:lowsupply_price}
p^*(B)=\ln\left(1+\frac{G}{B}\right)-\frac{G}{B+G}.
\end{equation}
\item \emph{Monopolist's high investment regime:} if $B>B_{th}$ (e.g., $B^H$ in Fig.
\ref{fig:DintersectsS}), then it is optimal to choose a price such
that supply exceeds demand,
\begin{equation}\label{eq:highsupply_price}
p^*(B)=0.468.
\end{equation}
\end{itemize}

\subsubsection{Main proof of duopoly symmetric pricing structure}
Now let us consider the two operator case based on the monopolist's
result. Suppose that there exists an equilibrium ($p^*_i,p^*_j$)
with $p_i^*\neq p_j^*$. The operator announcing lower price acts as
a monopolist. Without loss of generality, we assume that $B_i\leq
B_j$. In the following analysis, we examine asymmetric pricing
equilibrium in all ($B_i,B_j$) possibilities.
\begin{itemize}
\item[(a)] If $B_{th}<B_i\leq B_j$, then any operator will not
announce their prices higher than 0.468, in a fear of losing all its
realized demand to its competitor. And the operator with the lower
price will always increase its price to infinitely approach the
other operator's price, and thus no equilibrium can be obtained.
\item[(b)] If $B_i\leq B_{th}\leq B_j$, then operator $i$ will
announce $p_i^*\leq0.468$ and operator $j$ will announce $p_j^*\leq
\ln(1+G/B_i)-G/(B_i+G)$. And the operator with the lower price will
always increase its price to infinitely approach the other
operator's price, and no equilibrium can be obtained.
\item[(c)] If $B_i\leq B_j<B_{th}$, then operator $i$ will announce
$p_i^*\leq \ln(1+G/B_j)-G/(B_j+G)$, and operator $j$ will announce
$p_j^*\leq \ln(1+G/B_i)-G/(B_i+G)$. And the operator with lower
price will always increase its price to infinitely approach the
other operator's price, and thus no equilibrium can be obtained.
\end{itemize}
Thus there only exists possible symmetric pricing equilibrium.

\subsubsection{Main proof of duopoly pricing equilibrium}
Now we consider the duopoly pricing equilibrium ($p_i^*,p_j^*$)
which should satisfy $p_i^*=p_j^*$. Let us write symmetric price as
$p^*$. Since a user $k$'s demand is $g_k/H(p^*)$ in
(\ref{eq:userdemand_generalSNR}), the users' total preferred demand
is $G/H(p^*)$. Following a similar analysis in Section
\ref{subsec:stageIII}, the realized demands of the two operators are
$$Q_i=\min\left(B_{i},
\frac{G}{2H(p^*)}+\max\left(\frac{G}{2H(p^*)}-B_{j},0\right)\right),$$
$$Q_j=\min\left(B_{j},\frac{G}{2H(p^*)}+\max\left(\frac{G}{2H(p^*)}-B_{i},0\right)\right).$$
Two operators' revenues are $R_i=p^*Q_i$ and $R_j=p^*Q_j$,
respectively. Assume, without loss of generality, that $B_i\leq
B_j$. In the following analysis, we examine all ($B_i,B_j$)
possibilities, and check if there exists any symmetric pricing
equilibrium.

\begin{itemize}
\item[(a)] If $B_{th}<B_i\leq B_j$, we have shown that $p^*\leq
0.468$. We investigate possible pricing equilibrium given operator
$j$'s investment amount.
\begin{itemize}
\item If $p^*\leq\ln(1+G/B_j)-G/(B_j+G)$, operator $j$'s investment is not enough to satisfy the total preferred demand.
\begin{itemize}
\item If $B_i\leq \frac{G}{2H(p^*)}$, operator $i$ can not realize even all its preferred demand.
\begin{itemize}
\item If total supply is larger than total preferred demand (i.e.,
$B_i+B_j>G/H(p^*)$), then operator $j$ will slightly decrease its
price to attract much more preferred demand.
\item If total supply is smaller than total preferred demand (i.e., $B_i+B_j\leq
G/H(p^*)$), operator $j$ will slightly increase its price while its
attracted preferred demand will not change.
\end{itemize}
\item If $B_i>\frac{G}{2H(p^*)}$, operator $i$'s investment is enough to cover its preferred demand.
Then operator $j$ will slightly decrease its price to attract much
more preferred demand from its competitor.
\end{itemize}
\item If $p^*>\ln(1+G/B_j)-G/(B_j+G)$, operator $j$'s investment
amount is enough to satisfy the total preferred demand. Then
operator $j$ will slightly decrease its price to attract much more
preferred demand.
\end{itemize}
Hence, there does not exist an equilibrium in case (a).

\item[(b)] If $B_i\leq B_{th}\leq B_j$, we have shown that
$p^*\leq0.468$ and $p^*\leq\ln(1+G/B_i)-G/(B_i+G)$. Then we can show
that there does not exist an equilibrium in case (b) by a similar
argument as in case (a).

\item[(c)] If $B_i\leq B_j<B_{th}$, we will first show that total
supply equals total preferred demand at any possible equilibrium
(i.e., $p^*=\ln(1+G/(B_i+B_j))-G/(B_i+B_j+G)$).
\begin{itemize}
\item Suppose that at an equilibrium
$p^*<\ln(1+G/(B_i+B_j))-G/(B_i+B_j+G)$ and thus total supply is
smaller than total preferred demand. Then operator $j$ will slightly
increase its price while its attracted preferred demand will not
change.
\item Suppose that at an equilibrium
$p^*>\ln(1+G/(B_i+B_j))-G/(B_i+B_j+G)$ and thus total supply is
larger than total preferred demand. Then operator $j$ will slightly
decrease its price to attract much more preferred demand.
\end{itemize}
Thus total supply equals total preferred demand with
$p^*=\ln(1+G/(B_i+B_j))-G/(B_i+B_j+G)$ for possible equilibrium. Let
us check if this is indeed an equilibrium given total investment
amount.
\begin{itemize}
\item If $B_i+B_j>B_{th}$, then we have $p^*<0.468$. Operator
$i$ gets its supply sold out, and essentially it acts as a
monopolist in serving its realized users in monopolist's high
investment regime. Thus operator $i$ will increase its price to
$0.468$.

\item If $B_i+B_j\leq B_{th}$, then we have $p^*\geq 0.468$.
Operator $i$ gets its supply sold out, and essentially acts as a
monopolist in serving its realized users in monopolist's low
investment regime. Thus it's optimal to stick with current price.
Similarly, operator $j$ will also stick with current price.
\end{itemize}
Thus there exists a unique pricing equilibrium with
$p_i^*=p_j^*=\ln(1+G/(B_i+B_j))-G/(B_i+B_j+G)$ for low investment
regime (i.e., $B_i+B_j\leq B_{th}$) only.
\end{itemize}\hfill$\rule{2mm}{2mm}$

\subsection{Proof of Theorem \ref{thm:ObgeneralSNR}}\label{proofObservations}
\subsubsection{Proof of Observation \ref{ob:linearinvestment}}
The competition between two operators in Stage I can be modeled as
the following investment game:
\begin{itemize}
\item Players: operators $i$ and $j$.
\item Strategy space: two operators will choose ($B_i,B_j$) from the
set $\mathcal{B}=\{(B_i,B_j):B_i+B_j\leq B_{th}\}$.
\item Payoff function: two operators want to maximize their own profits
$\pi_i(B_i,B_j)$ in (\ref{eq:iprofit_pricing}) and $\pi_j(B_i,B_j)$
in (\ref{eq:jprofit_pricing}).
\end{itemize}
The best response of operator $i$ (i.e., $B_i^*(B_j)$) also equals
$\arg\max_{0\leq B_i\leq B_{th}-B_j}\pi_i(B_i,B_j)/G$. Notice that
\begin{multline}\pi_i(B_i,B_j)/G\\=\frac{B_i}{G}\left[\ln\left(1+\frac{1}{\frac{B_i}{G}+\frac{B_j}{G}}\right)-\frac{1}{\frac{B_i}{G}+\frac{B_j}{G}+1}-C_i\right],\nonumber\end{multline}
where $B_i$ always appears together with $G$. Thus $B_i^*(B_j)$ is
linear in $G$ and we can similarly show that $B_j^*(B_i)$ is also
linear in $G$. Since the possible equilibrium $(B_i^*,B_j^*)$ is
derived by joint solving equations $B_i^*=B_i^*(B_j^*)$ and
$B_j^*=B_j^*(B_i^*)$, $B_i^*$ and $B_j^*$ are both linear in the
users' aggregate wireless characteristics $G$.

\subsubsection{Proof of Observation \ref{ob:indepprice}}
It is obvious that the symmetric pricing equilibrium in
(\ref{eq:priceGeneralSNR})
is determined by $B_i^*/G$ and $B_j^*/G$ only. Since we have shown
that operators' equilibrium investment decisions are both linearly
proportional to $G$, thus the pricing equilibrium
$p_i^*(B_i^*,B_j^*)=p_j^*(B_i^*,B_j^*)$ is independent of the users'
wireless characteristics.

\subsubsection{Proof of Observation \ref{ob:demandSNR}}
Assume, without loss of generality, that a user $k$ purchases
bandwidth from operator $i$. We have shown in
(\ref{eq:userdemand_generalSNR}) that each user $k$'s equilibrium
demand is always positive, linear in $g_k$. And it is also
decreasing in symmetric equilibrium price since positive function
$H(p_i^*)$ is increasing in $p_i^*$.
%

And the SNR of user $k$ is $\mathtt{SNR}_k=H(p_i^*)$, which also
equals $H(p_j^*)$ at symmetric pricing equilibrium. Thus each user
$k\in\mathcal{K}$ achieves the same SNR independent of its wireless
characteristic $g_k$. And it is clear that user $k$'s payoff in
(\ref{eq:payoff_generalSNR}) is independent of user $k$'s wireless
characteristic $g_k$.

\subsubsection{Proof of Observation \ref{ob:SNRCiCj}}
It is clear that as duopoly's symmetric price increases, the users'
achieved SNR increases but their payoffs decrease in general SNR
regime. To prove Observation \ref{ob:SNRCiCj}, we only need to show
that as leasing cost $C_i$ or $C_j$ increases, equilibrium price
will also increase. In the following analysis, we first show that
symmetric equilibrium price increases as $B_i$ or $B_j$ decreases.
Then we show that $B_i$ or $B_j$ decreases as leasing cost $C_i$ or
$C_j$ increases.

\begin{itemize}
\item It is easy to check that the first derivatives of $p_i^*(B_i,B_j)$ over $B_i$ and $B_j$
are both negative,
Thus duopoly's symmetric equilibrium price increases as $B_i$ or
$B_j$ decreases.

\item Operator $i$'s
revenue is $R_i(B_i,B_j)=B_ip_i^*(B_i,B_j)$ with $p_i^*(B_i,B_j)$ in
(\ref{eq:priceGeneralSNR}), and its profit in
(\ref{eq:iprofit_pricing}) also equals
$\pi_i(B_i,B_j)=R_i(B_i,B_j)-B_iC_i$. Due to the strict concavity of
$\pi_i(B_i,B_j)$ over $B_i$, operator $i$ will optimally lease
$B_i^*$ to make $\partial \pi_i(B_i,B_j^*)/\partial
B_i\mid_{B_i=B_i^*}=0$ (i.e., $\partial R_i(B_i,B_j^*)/\partial
B_i\mid_{B_i=B_i^*}=C_i$). And $R_i(B_i,B_j^*)$ is concave in $B_i$
by checking the twice derivative of $R_i(B_i,B_j)$ over $B_i$,
\begin{multline}\frac{\partial^2 R_i(B_i,B_j^*)}{\partial
B_i^2}=-\frac{G^2}{(B_i+B_j^*+G)^3(B_i+B_j^*)^2}\cdot\\
[-B_i^2+GB_i+B_iB_j^*+2GB_j^*+2B_j^{*2}],\nonumber\end{multline}which
is negative due to $B_i+B_j\leq B_{th}$. Thus $B_i^*$ decreases as
$C_i$ increases.
\end{itemize}

Hence, users' equilibrium SNR increases with the costs $C_i$ and
$C_j$, and their payoffs decrease with the
costs.\hfill$\rule{2mm}{2mm}$

\subsection{Proof of Theorem \ref{thm:Global_price}}\label{Proof_Thm5}

If $B_i>0$ and $B_j=0$ in the coordinated case, only operator $i$
will then participate in pricing stage and it becomes a monopolist.
According to \cite{report_Infocom}, the optimal leasing amount of
monopolist is in low supply regime and the optimal price is to make
the users' total demand equal to its supply. Thus operator $i$ will
announce the unique price
$$p_i^{co}=\ln\left(\frac{G}{B_i}\right)-1.$$ Similar result can
be obtained for $B_i=0$ and $B_j>0.$

If $\min(B_i,B_j)>0$, both the coordinated operators will
participate in the pricing stage. Since the duopoly's payments
($B_iC_i$ and $B_jC_j$) are already determined, the two operators
will cooperate to maximize their total revenue only in pricing
stage. Without loss of generality, we assume $p_i^{co}\leq
p_j^{co}$, and find the optimal pricing strategies of coordinated
duopoly as follows.

\begin{itemize}
\item We first show the feasible range of $p_i^{co}$ and $p_j^{co}$ should be $p_i^{co}\leq \ln\left(\frac{G}{B_i+B_j}\right)-1\leq
p_j^{co}.$ The reason is as follows. According to Proposition 1,
duopoly will set the prices such that total supply equals to the
users' total demand and it is easy to check that
$$B_ie^{1+p_i^{co}}+B_je^{1+p_j^{co}}=G.$$ Then we conclude that
$(B_i+B_j)e^{1+p_i^{co}}\leq G\leq (B_i+B_j)e^{1+p_j^{co}},$ and
thus $p_i^{co}\leq \ln\left(\frac{G}{B_i+B_j}\right)-1\leq
p_j^{co}.$

\item Then we derive the relation between $p_i^{co}$ and $p_j^{co}$.
All the users have priority to purchase bandwidth from operator $i$
who charges less, and operator $i$'s revenue is
$p_i^{co}\min(B_i,Ge^{-(1+p_i^{co})})$. We then discuss different
relation between operator $i$'s supply and preferred demand.

\begin{itemize}
\item If $B_i> Ge^{-(1+p_i^{co})}$, then operator $i$'s supply is
excessive compared with the preferred demand. Due to Proposition 1,
this will not happen in coordinated case.
\item If $B_i\leq Ge^{-(1+p_i^{co})}$, then operator $i$'s supply is
not enough to meet the preferred demand and we have $B_i=
\sum_{k\in\mathcal{K}_i^R}g_ke^{-(1+p_i^{co})}$. And the left demand
going to operator $j$ will be
$(G-\sum_{k\in\mathcal{K}_i^R}g_k)e^{-(1+p_j^{co})}$. Since
Proposition 1 requires total demand equals total supply, the
operator $j$ should decide a price such that its supply equals the
left demand, i.e.,
$p_j^{co}=\ln\left((G-B_ie^{1+p_i^{co}})/{B_j}\right)-1.$
\end{itemize}
Thus we conclude that $p_j^{co}$ is a function of $p_i^{co}$ and
rewrite it as
$$p_j^{co}(p_i^{co})=\ln\left(\frac{G-B_ie^{1+p_i^{co}}}{B_j}\right)-1.$$

\item In pricing stage, the total profit maximization problem in coordinated case is
equivalent to the total revenue maximization problem. And the total
revenue can be expressed as a function of $p_i$ only. The solution
to the total revenue maximization problem is
$$p_i^{co}=\arg\max_{0\leq p_i\leq \ln\left(\frac{G}{B_i+B_j}\right)-1}B_ip_i+B_jp_j^{co}(p_i^{co}).$$
Since $B_ip_i+B_jp_j^{co}(p_i^{co})$ is an increasing function of
$p_i$ on range $0\leq p_i\leq \ln\left(\frac{G}{B_i+B_j}\right)-1$,
we have
$$p_i^{co}=\ln\left(\frac{G}{B_i+B_j}\right)-1.$$ By substituting
the value of $p_i^{co}$ into
$B_ie^{1+p_i^{co}}+B_je^{1+p_j^{co}}=G,$ we also obtain
$$p_j^{co}=\ln\left(\frac{G}{B_i+B_j}\right)-1.$$

Hence, the optimal pricing strategies of coordinated duopoly are
$$p_i^{co}=p_j^{co}=\ln\left(\frac{G}{B_i+B_j}\right)-1.$$
\end{itemize}
\hfill$\rule{2mm}{2mm}$


\end{document}